\documentclass[aps,pre,reprint,twocolumn,showpacs,superscriptaddress]{revtex4-2}
\usepackage{epsfig}
\usepackage{natbib}
\usepackage[normalem]{ulem}
\usepackage{xspace}
\usepackage{xcolor}
\usepackage{graphicx}
\usepackage{amssymb}
\usepackage{amsmath}
\usepackage{stackrel}
\usepackage{appendix}
\usepackage{url}
 \usepackage{enumitem}
 \usepackage{hyperref}
\usepackage{bm}
\usepackage{mathrsfs}
\usepackage{booktabs}

\hypersetup{
	colorlinks=true,
	linkcolor=blue,
	citecolor=blue,
	urlcolor=blue
}

\begin{document}

\title{Dynamical phase transition for a logarithm observable in chaotic systems}

\author{Lucianno Defaveri}
\affiliation{Racah Institute of Physics, Hebrew University of Jerusalem, Jerusalem 91904, Israel}
\author{Naftali R. Smith}
\email{naftalismith@gmail.com}
\affiliation{Racah Institute of Physics, Hebrew University of Jerusalem, Jerusalem 91904, Israel}

\begin{abstract}
We consider the full distribution $P_N(A)$ of the dynamical observable
$A = - \sum_{i=1}^N \ln x_i$
where $x_1, \dots, x_N$ is a sequence generated by a chaotic map $f(x)$ via $x_{n+1} = f(x_n)$.  We study the particular cases where $f(x)$ is the doubling, tent or logistic map, and focus on the limit $N \gg 1$.
Typical fluctuations of $A$ are Gaussian with mean $\left\langle A\right\rangle$ and variance which are both of order $\sim N$.
We find that the left tail of the distribution ($A < \left\langle A\right\rangle$) scales as $P_{N}(A) \sim e^{-NI(A/N)}$, and calculate the rate function $I(a)$ which depends on the map $f(x)$.
Interestingly, we find that the right tail of the distribution ($A > \left\langle A\right\rangle$) follows an anomalous scaling:
$P_{N}(A) \sim e^{-N^{1/3} F((A-\left\langle A\right\rangle) / N^{2/3})}$.
The large-deviation function $F(y)$ has a universal form for the three maps, differing only in the scaling constants. At a critical value $y=y_c$, it exhibits a corner singularity which has the character of a condensation transition. The transition separates between a subcritical phase $0< y<y_c$ in which the fluctuations are Gaussian and arise homogeneously from all of the $x_i$'s, and a supercritical phase $y>y_c$ in which a small fraction of the $x_i$'s give a macroscopic contribution to the fluctuation.
Despite this anomalous behavior of the right tail, we find that the cumulants of the distribution $P_N(A)$ grow linearly with $N$ at large $N$, and calculate the associated coefficients.
\end{abstract}

\maketitle

\section{Introduction}

\subsection{Background}

Large deviations -- or rare events -- can have dramatic, or even catastrophic consequences: Examples include heatwaves, earthquakes, pandemic outbreaks, and stock-market crashes. It is therefore important to understand how they occur, and to be able to quantify their probabilities.
Large deviations are also of fundamental importance in statistical mechanics, since they provide a means of studying systems that are out of equilibrium. As a result, they have been extensively studied over recent decades \cite{DZ1998, Hollander2000, Touchette2009, MS2017, Nickelsen2018}, and remarkable behaviors have been uncovered. These include singularities in large-deviation rate functions \cite{Baek15, Baek17, Baek18, Agranov2020, Smith2022, ACJ22, ACJ23, DBK24, DBK25}, which are usually interpreted as dynamical phase transitions, in analogy with systems in thermal equilibrium.

Much of the existing work has focused on stochastic processes. These include Brownian motion \cite{Agranov2020, Smith24Absx, BMT23}, lattice gases \cite{ACJ22, MFTreview}, stochastic surface growth \cite{Janas2016, SMS2018, KLD18, MKV}, active particles \cite{MLS24, Smith23NESS, SBS20}, but this list is far from exhaustive. In these cases, the randomness -- which ultimately leads to the occurrence of the rare event of interest -- usually originates in the intrinsic stochasticity of the process, while the initial configuration plays a less important role.
This is particularly true when studying large deviations of long-time averages in stochastic processes (also known as ``dynamical observables''): Here the initial condition is quickly ``forgotten'', and its effect on the observable in question is negligible.

Large deviations in deterministic, chaotic systems \cite{Grassberger1985,Prasad1999,Anteneodo2004,Nakao2006,Lai2007}  have attracted considerable renewed interest over the last few years \cite{Smith2022, Gutierrez2023, Monthus2024, Lippolis2024, Monthus2025, Gonzalez-Koda2025, DS26}. One motivation for studying these systems comes from weather and climate systems, which are often modeled as deterministic and chaotic (rather than stochastic).
Here the situation is very different to the case of stochastic processes: There is no intrinsic randomness in the dynamics, and instead, all of the randomness comes from the choice of initial conditions. This difference can lead to behaviors that are qualitatively different to what is observed in stochastic systems. For instance, as we showed in a recent work, for a certain class of dynamical observables, the statistics do not obey the standard central limit theorem nor the large-deviations principle \cite{DS26}, in a manner that is not known to occur in stochastic dynamics.

It is therefore natural to ask in what other respects large deviations in chaotic systems may or may not differ from the more extensively studied stochastic systems.
One particular phenomenon that is known to occur in large deviations of stochastic processes, but has not yet been demonstrated in deterministic, chaotic systems, is known as the condensation transition.
This type of transition occurs in a broad range of stochastic processes, including the Ornstein-Uhlenbeck process \cite{Nickelsen2018,  Smith22OU, NT22, Ferre25, BVBB25} (and some of its generalizations \cite{MeersonGaussian19,VM25}), stochastic resetting \cite{DH2019, SmithMajumdar22}, reflected Brownian motion \cite{BH20,Smith24Absx} and active particles \cite{GM19,MLMS21,MGM21}. In fact, the condensation transition occurs in the far simpler case of sums of i.i.d. random variables whose distribution tails decay sub-exponentially \cite{BKLP20, SmithCondensation25, BH26, Burov26}. It has been observed to occur for sums of correlated random variables too: In particular, for linear spectral statistics of random matrices \cite{VMS24}.
Condensation transitions are often associated with an anomalous large-deviations scaling. They are also related to the ``big-jump'' principle. One way to describe this principle is that for certain types of systems, the large deviation distribution tails of long-time averages are dominated by a single localized event which dominates the fluctuation from the expected behavior.

The goal of this paper is to show that such condensation transitions, accompanied by anomalous large-deviations scalings, occur in chaotic systems as well. We achieve this by demonstrating these behaviors for particular examples which involve discrete-time, one-dimensional chaotic dynamics generated by the doubling, tent, or logistic maps.
In fact, we uncover behaviors that coincide, up to scaling factors, with those which have been observed previously in the stochastic systems described above, thus indicating that the condensation transition is universal to a broad class of systems. We find that the temporal correlations of the dynamics strongly affect the fluctuations when comparing to the i.i.d. case. Furthermore, we show that despite the anomalous scaling behavior of the fluctuations, the cumulants of the associated distribution exhibit the regular scaling, i.e., they grow linearly with the duration of the dynamics at long times.

\subsection{Model, observable, scaling behaviors and outline of the paper}

To facilitate the theoretical analysis, we focus in this work on discrete-time chaotic dynamics.
Following the general framework of Ref.~\cite{Smith2022}, we consider a sequence of values $\bm{x}_1, \bm{x}_2, \ldots, \bm{x}_N$, where $\bm{x}_n \in \mathbb{R}^d$, which is generated by a chaotic map $f: \mathbb{R}^d \to \mathbb{R}^d$ through
\begin{equation}
    \bm{x}_{n+1} = f(\bm{x}_n) \, . \label{eq:interative-map}
\end{equation} 
The initial condition $\bm{x}_1$ is sampled from a probability density function (PDF) which, for simplicity, we choose to be the steady-state distribution, or invariant measure (IM), $p_s(\bm{x}_1)$ of the dynamics.
We are interested in the distribution $P_N(A)$ of dynamical observables \cite{Touchette2009, Nickelsen2018, Smith2022, Gutierrez2023, Monthus2024}
\begin{equation}
\label{Adef}
    A = \sum_{n=1}^N g(\bm{x}_n) \, ,
\end{equation}
where $g:\mathbb{R}^d \to \mathbb{R}$,
so that $A/N$ is the time average of $g$ applied to the elements of the chaotic sequence. In particular, we are interested in the behavior of $P_N(A)$ in the limit $N \gg 1$, where universal behavior is expected to emerge.

The ``usual'' behavior of $P_N(A)$, i.e., for standard choices of the map $f$ and the function $g$ in Eq.~\eqref{Adef} is as follows. Typical fluctuations of $A$ obey a central limit theorem, with a mean and variance that can be calculated straightforwardly from the steady-state distribution and the two-time correlation function (see details below). Large fluctuations of $A$ from its typical value are described by the large-deviations principle (LDP)
\begin{equation}
\label{LDP}
    P_N(A) \sim e^{-NI(A/N)}\,,
\end{equation}
describing an exponential decay with respect to $N$, with a ``rate function'' $I(a)$. The minimum of the rate function $a_*$ is at the typical (or mean) value of $A/N$, and there $I(a_*) = 0$.

In this work, we will focus on the particular case of one-dimensional maps ($d=1$) from the unit interval $[0,1]$ to itself. We focus mostly on the particular cases of the doubling map $f(x) = 2x \! \mod \! 1$, the tent map $f(x) = 1 - 2\left|x-1/2\right|$ and the logistic map $f(x) = 4x\left(1-x\right)$.
Our main results were verified for all three maps; more general derivations and figures can be found in the Appendices. For simplicity, in the remainder of the manuscript, the figures and examples refer to the doubling map unless stated otherwise.
Moreover, we will consider the specific observable defined by the  function
\begin{align}
    g(x) = - \ln \, x \, ,
\end{align}
so that the observable $A$ is minus the logarithm of the product of the elements in the sequence, $A \! = \! -\ln \prod_{i=1}^N x_i$.
This choice of observable [and in particular, the fact that $g(x)$ is unbounded from above] enables the ``big jump'' principle to occur in the right tail of the distribution $P_N(A)$, and thus, the LDP \eqref{LDP} breaks down in the right tail. For a certain class of unbounded observables which includes our particular choice, the breaking down of the LDP has been rigorously proven in the mathematical literature \cite{NT20, Pontes26}. Moreover, bounds for the distribution tails were obtained, showing stretched exponential decay with $N$, in contrast to the exponential decay  \eqref{LDP} that occurs for bounded observables.
However, more precise estimates of the distribution $P_N(A)$, and/or a full description of the distribution that includes the typical fluctuations and tail regimes as well as the intermediate regime between them is currently lacking. The main goal of the current work is to provide such a description.
Some of our general derivations below are valid for a general $g(x)$, and we will  plug in $g(x) = - \ln x$ when we wish to specialize to this particular case.

The structure of the rest of the paper is as follows.
In Section \ref{sec:TypicalAndDV} we show that the typical fluctuations are Gaussian, and that the left tail ($A< \langle A \rangle$) obeys the LDP \eqref{LDP}, and calculate the associated rate function $I(a)$. We also show that the LDP \eqref{LDP} also describes the near right tail of the distribution. Moreover, we show that the cumulants of the distribution $P_N(A)$ grow proportionally to $N$ in the large $N$ limit and calculate the associated coefficients of proportionality.
In Section \ref{sec:condensation} we study the right tail ($A> \langle A \rangle$). We show that this tail obeys an anomalous LDP scaling, calculate the associated large-deviations function exactly, and show that it exhibits a corner singularity corresponding to a critical value $A=A_c$.
In Section \ref{sec:discussion} we summarize and discuss our main findings.
Some technical details of the theoretical derivations and of the numerical simulations are given in the Appendices.

\section{Typical fluctuations $A \simeq \langle A \rangle$ and Donsker-Varadhan theory for the left tail $A< \langle A \rangle$ and for the near right tail}
\label{sec:TypicalAndDV}

\subsection{Typical fluctuations}

Typical fluctuations of $A$ are Gaussian with a mean and variance that are relatively straightforward to calculate.
The mean of $A$ is simply given by
\begin{equation}
\label{meanA}
    \langle A \rangle =\sum_{i=1}^{N}\langle g(x_{i})\rangle =N\langle g(x_{1})\rangle =N\int g(x)p_{s}(x)dx \, .
\end{equation}
This equation is exact under the assumption that $x_1$ is sampled from $p_s(x)$, and otherwise it only holds approximately, in the limit $N \gg 1$.

The variance of $A$ is also proportional to $N$ at $N \gg 1$, and the constant of proportionality is given by the Green-Kubo formula. Explicitly, one has
\begin{equation}
\text{Var}(A)\simeq N\sum_{\delta=-\infty}^{\infty}\left[\left\langle g(\bm{x}_{0})g(\bm{x}_{\delta})\right\rangle _{S}-\left\langle g(\bm{x})\right\rangle _{S}^{2}\right]\,, \label{eq:variance-green-kubo}
\end{equation}
where $\left\langle \cdots\right\rangle _{S}$ denotes ensemble averaging, in which each of the $\bm{x}_i$'s is distributed according to the IM.
For the particular case where $f(x)$ is the doubling map, $f(x) = 2x \mod 1$ and $g(x) = -\ln x$, using $p_{s}(x)=1$ we find $\left\langle A\right\rangle =N$, while the variance is given by $\text{Var}(A)\simeq 3.71\,N$, where the prefactor is found by a numerical computation of the sum in \eqref{eq:variance-green-kubo}.
Remarkably, for the logistic map, the sum in Eq.~\eqref{eq:variance-green-kubo} can be evaluated in closed form and yields exactly $\mathrm{Var}(A)\simeq \pi^{2} N$. This follows from the conjugation of the logistic map to the tent map, under which the observable admits an explicit discrete Fourier series expansion and the correlations in Eq.~\eqref{eq:variance-green-kubo} can be computed exactly, see Appendix~\ref{app:exact-green-kubo-logistic}. For all three maps, these values of the variance agree with the second cumulant $\lambda_2$ obtained from the tilted operator via a power method, and with direct trajectory simulations, whose sample variance approaches $\lambda_2 N$ asymptotically at large $N$. Details are found in Appendix~\ref{app:numerical-SCGF-power-method}.

From Eq.~\eqref{meanA} together with the fact that the typical fluctuations scale as $A - \langle A \rangle \sim \sqrt{N}$ it follows that, in the limit $N\to\infty$, the time average $A/N$ of $g(x_i)$ coincides with the ensemble average, as follows from ergodicity. Nevertheless, at large but finite $N \gg 1$, large fluctuations from the ensemble average are possible, and in most of the remainder of the paper, we aim to quantify their likelihood.

\subsection{The left tail: Donsker-Varadhan theory}

\begin{figure}
    \centering
    \includegraphics[width=\linewidth]{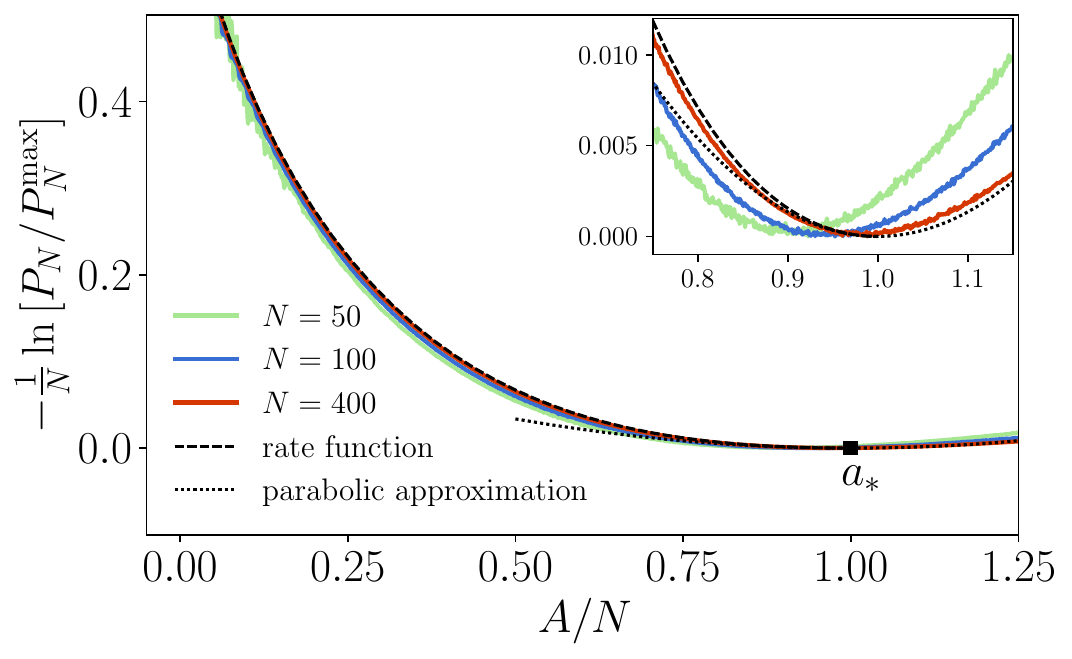}
    \caption{The theoretical rate function $I(a)$ for the doubling map (dashed line), plotted together with 
    $-(1/N)\ln\left[P_N(A)/P_N^{\mathrm{max}}\right]$ as a function of $a=A/N$, for $N=50$, $100$ and $400$ (solid lines), obtained from simulations (see Appendix~\ref{app-sec:importance-sampling}). As $N$ grows, the numerical curves converge toward the rate function $I(a)$ of Eq.~\eqref{eq:legendre-fenchel}, computed from the SCGF which, in turn, is obtained numerically using a power method (see Appendix~\ref{app:numerical-SCGF-power-method}). The dotted line shows the parabolic approximation $I(a)\simeq I_2(a-a_*)^2$, with $I_2 \simeq 0.135$, around the minimum $a_*=1$ (marked with a square), which describes the Gaussian regime of typical fluctuations. Inset: zoom on the region around $a_*$. We highlight that even to the right of $a_*$, despite the large deviation theory breaking down, it is still possible to observe convergence to the parabolic approximation (dotted lines).}
    \label{fig:rate-function-left}
\end{figure}

The left tail of the distribution $P_N(A)$ is described by the standard LDP \eqref{LDP}, since $g(x) = -\ln x$ is bounded from below. The rate function $I(a)$ is formally defined as \cite{Touchette2009,Smith2022,Gutierrez2023,Monthus2024}
\begin{equation}
    I(a \equiv A/N) = - \lim_{N \to \infty} \frac{\ln P_N(A)}{N} \, . \label{eq:Ia-from-PN}
\end{equation}
The function $I(a)$ is convex and its minimum is at the ensemble-average value $a_* = \langle g(x)\rangle_S$ ($=1$ for the doubling map), where $I(a_*) = 0$. Note that $I(a)$ only describes the left tail of the distribution, $a\le a_*$. For $a>a_*$, as we will show in the next section, the LDP \eqref{LDP} breaks down, and the rate function $I(a)$ is identically zero there: $I(a > a_*) \equiv 0$, thus indicating that the large-deviation scaling behavior is different from \eqref{LDP}.
Near the minimal point, $a \simeq a_*$, $I(a)$ assumes a parabolic approximation $I(a) \simeq I_2(a-a_*)^2$, which matches smoothly with the Gaussian distribution of typical fluctuations \cite{Smith2022}. This is illustrated in Fig.~\ref{fig:rate-function-left}, where the numerical computations of
$-(1/N)\ln\left[P_N(A)/P_N^{\mathrm{max}}\right]$, where $P_{N}^{\mathrm{max}}=\max_{A}P_{N}\left(A\right)$,
at increasing values of $N$ are seen to converge toward the rate function $I(a)$, and the parabolic approximation is seen to hold in a neighborhood of $a_*$.

We will now briefly explain how to calculate the rate function $I(a)$ (for a more general and detailed derivation, see e.g.~\cite{Smith2022}).
It is useful to first calculate the scaled cumulant generating function (SCGF) $\lambda(k)$, defined as
\begin{equation}
    \left\langle e^{k A} \right\rangle \sim e^{ N \lambda(k)} \, . \label{eq:SCGF}
\end{equation}
The Legendre-Fenchel transform of the SCGF then yields the rate function \cite{Grassberger1985,Touchette2009,Smith2022,Gutierrez2023,Monthus2024},
\begin{equation}
    I(a) = \sup_{k \in \mathbb{R}} \left[ k a - \lambda(k) \right] \, . \label{eq:legendre-fenchel}
\end{equation}

To calculate the SCGF $\lambda(k)$, we use the G\"{a}rtner-Ellis theorem, which states that $\lambda(k)$ is given by the logarithm of the largest (real) eigenvalue of an operator $L_k$, called the ``tilted generator'' of the dynamics. $L_k$ is given by \cite{Touchette2009,Laffargue2013,Smith2022,Gutierrez2023,Monthus2024}
\begin{equation}
L_{k}\psi_{R}(x)=\sum_{z\in f^{-1}(x)}\frac{e^{kg(z)}\psi_{R}(z)}{\left|f'(z)\right|} \, . \label{eq:tilted-operator-definition}
\end{equation}
Note that $L_{k=0}$ is simply the Frobenius-Perron operator, that gives the distribution of $f(x)$ from the distribution of $x$.
For our particular $f(x)$ and $g(x)$, the eigenvalue equation for the tilted operator reads
\begin{equation}
    L_{k}\psi_{R}(x)=\sum_{z\in\left\{ \frac{x}{2}, \frac{1 + x}{2}\right\} }\frac{z^{-k}\psi_{R}(z)}{2}=e^{\lambda(k)}\psi_{R}(x)\,.\label{eq:tilted-operator}
\end{equation}
In \eqref{eq:tilted-operator}, $\psi_R(x)$ and $e^{\lambda(k)}$ are the right eigenfunction and eigenvalue of $L_k$, respectively.
The DV eigenvalue problem is not easy to solve analytically for general $f(x)$ and $g(x)$. This has been achieved in some particular cases \cite{Smith2022, Monthus2024, DS26}, but the methods used previously cannot be naturally extended to our case (except for special values of $k$ in the logistic map, as we show below).

Importantly, the DV eigenvalue problem is ill-defined for $k>0$, since the spectrum of $L_k$ is unbounded from above, 
$\lambda\left(k>0\right) = \infty$.
This divergence reflects the fact that fluctuations of $A$ above its mean are only sub-exponentially unlikely (as a function of $N$).
As we show in Sec.~\ref{sec:condensation}, for $A \gg \langle A \rangle$, the PDF develops a stretched-exponential tail, $P_N(A) \sim e^{- c \sqrt{A}}$, and this decay is too slow to be compensated by any $e^{kA}$ with $k>0$. Recalling the definition \eqref{eq:SCGF}, this indeed leads to $\lambda\left(k>0\right) = \infty$.

A similar situation, in which the tilted operator's spectrum is unbounded from above for some values of the tilting parameter, is known to occur when studying large deviations in several stochastic systems, including the Ornstein-Uhlenbeck and related processes \cite{Nickelsen2018,  MeersonGaussian19, Smith22OU, NT22, Ferre25, BVBB25, VM25}, fractional or standard Brownian motion with stochastic resetting \cite{DH2019, SmithMajumdar22}, Brownian motion with a drift towards a reflecting wall \cite{BH20,Smith24Absx}, and active particles \cite{GM19,MLMS21,MGM21}.

\begin{figure}
    \centering
    \includegraphics[width=0.95\linewidth]{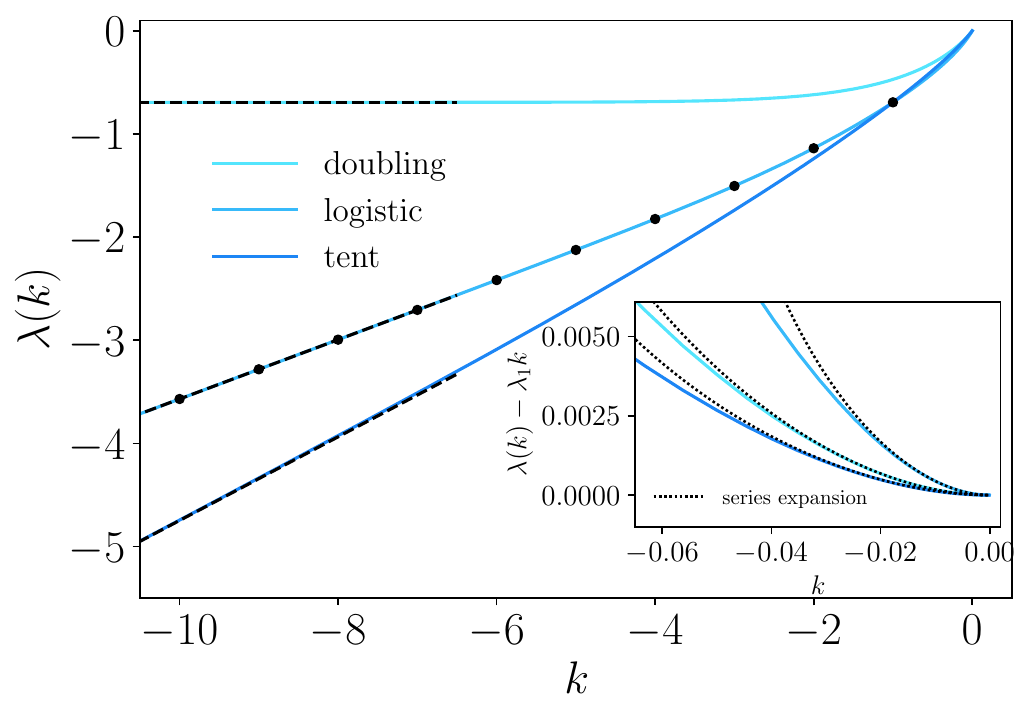}
    \caption{The SCGF $\lambda(k)$ for the doubling, tent and logistic maps, obtained by solving the eigenvalue problem in Eq.~\eqref{eq:tilted-operator} numerically using a power method (see Appendix~\ref{app:numerical-SCGF-power-method}). For the logistic map, the exact values at negative integer $k$ (black dots), obtained in Appendix~\ref{app:negative-integer-k}, are in excellent agreement with the numerical curve. The dashed black lines show the asymptotic behavior at $k \to -\infty$, $\lambda(k) \simeq k\,g(x_*) - \ln\left|f'(x_*)\right|$, which is dominated by trajectories that remain near the fixed point $x_*$ of the map at which $g(x) = -\ln x$ is smallest, namely $x_* = 1$, $2/3$ and $3/4$ for the doubling, tent and logistic maps, respectively, with $\left|f'(x_*)\right| = 2$ in all three cases. The inset shows $\lambda(k)- \lambda_1 k$ at small $|k|$, compared with the truncated series expansion \eqref{eq:SCGF-perturbative-expansion} with coefficients $\lambda_n$ from Table~\ref{tab:cumulants} (dotted lines), up to the fourth order, demonstrating that the series expansion around $k=0$ shows excellent agreement. Therefore, the rate function $I(a)$ can be approximated by a series expansion around $a_*$, with the higher-order coefficients (those beyond the parabolic approximation) obtained from the $\lambda_n$ as described in Appendix~\ref{app:DVsubleading}. The coefficients are found in Table~\ref{tab:rate-coefficients}.}
    \label{fig:scgf_all_maps}
\end{figure}

For $k\le 0$, we solve the eigenvalue problem \eqref{eq:tilted-operator} numerically using the power method \cite{Trefethen1997, Saad2011, Golub2013}, as detailed in Appendix \ref{app:numerical-SCGF-power-method}. The resulting $\lambda(k)$ is shown in Fig.~\ref{fig:scgf_all_maps} for the doubling, tent and logistic maps.
Interestingly, for the logistic map, $\lambda(k)$ can also be found analytically whenever $k$ is a negative integer. In this case, the eigenvalue is the largest real root of a polynomial of degree $-k$ whose coefficients can be found analytically. The first two values are 
\begin{equation}
\label{lambdam1m2}
    \lambda(-1) = -\ln 2, \quad \lambda(-2) = \ln\left(\frac{1+\sqrt{17}}{16}\right) \, .
\end{equation}
For $k \le -3$, in general the root of the polynomial is found numerically, for instance, we find that $\lambda(-3) \simeq -1.5043$ and $\lambda(-4) \simeq -1.8248$. The derivation is given in Appendix \ref{app:negative-integer-k}.

Given $\lambda(k)$, we obtain $I(a)$ by a numerical computation of the Legendre-Fenchel transform in Eq.~\eqref{eq:legendre-fenchel}, which reduces to the Legendre transform
\begin{equation}
    a = \lambda'(k), \qquad I(a) = ka-\lambda 
\end{equation}
since, as we found, $\lambda(k)$ is differentiable and convex (for $k < 0$). 
For the doubling map, the result is plotted in Fig.~\ref{fig:rate-function-left} and shows good agreement with the left tail of $P_N(A)$ for $N=50$, which we obtained numerically. The agreement improves as $N$ is increased, as can be seen in the figure.

In addition to yielding the rate function via the Legendre-Fenchel transform, the SCGF governs the growth of the cumulants of $A$ with $N$. Within DV theory, we expect all cumulants grow linearly with $N$ at $N \gg 1$, that is,
\begin{equation}
    \left\langle A^{n}\right\rangle _{c}\simeq\lambda_{n}N \, , \label{eq:cumulants-scale-with-N}
\end{equation}
 where the coefficients $\lambda_n$ are given by the series expansion of the SCGF around $k=0$, explicitly,
\begin{equation}
\lambda\left(k\right)=\sum_{n=1}^{\infty}\frac{\lambda_{n}k^{n}}{n!}\,. \label{eq:SCGF-perturbative-expansion}
\end{equation}
Both properties are immediate consequences of Eq.~\eqref{eq:SCGF}.
The first two coefficients $\lambda_1$ and $\lambda_2$ describe the mean and variance, and can be shown to coincide with the formulas \eqref{meanA} and \eqref{eq:variance-green-kubo}, respectively \cite{Shibata2002, Touchette2009}.

\begin{figure}
    \centering
    \includegraphics[width=0.95\linewidth]{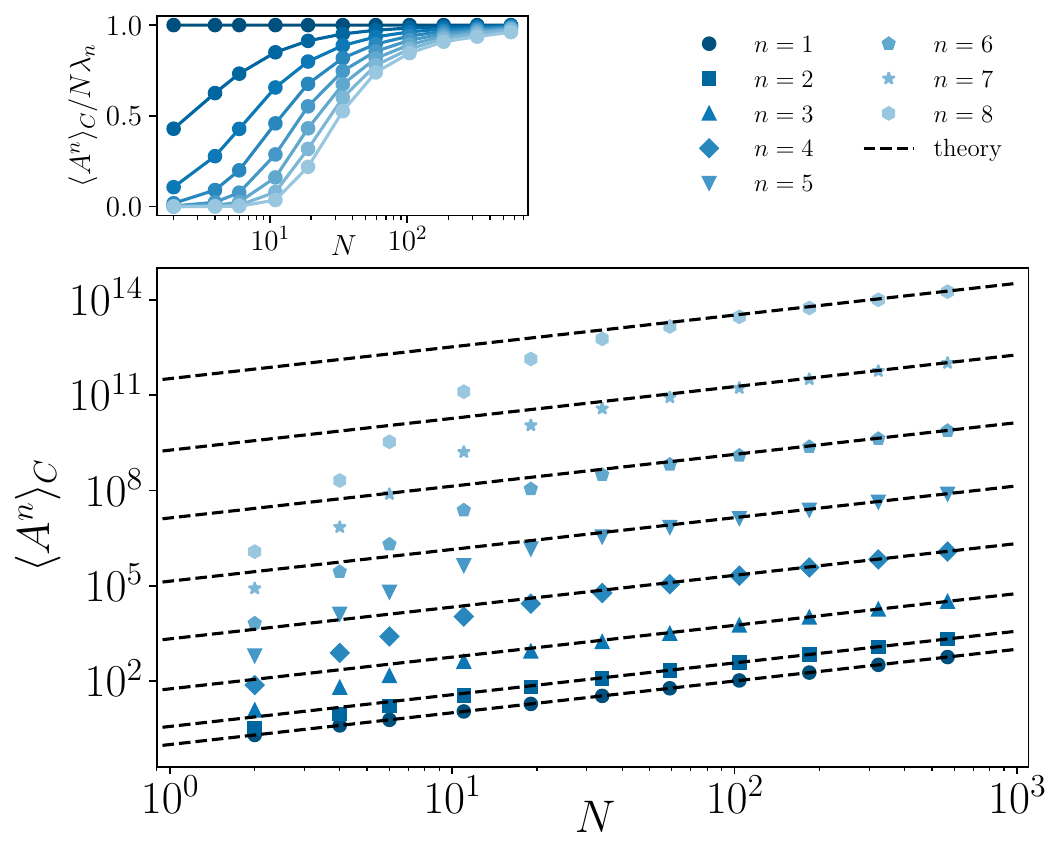}
    \caption{Cumulants $\langle A^n \rangle_c$ of the observable $A = -\sum_{i=1}^N \ln x_i$ for the doubling map. The dashed black lines represent the theory in Eq.~(\ref{eq:cumulants-scale-with-N}), with the values of $\lambda_n$ obtained through the tilted operator (see Appendix\,\ref{app:powerSeriesLambdak}).
    Top left panel: $\langle A^n \rangle_c / (\lambda_n N)$
    as a function of $N$ for orders $n = 1, \ldots, 8$, showing in a linear scale convergence to the linear-growth prediction of Eq.~(\ref{eq:cumulants-scale-with-N}), $\langle A^n \rangle_c \simeq \lambda_n N$. The specific values of $\lambda_n$ can be found in the Appendix, in Table~\ref{tab:cumulants}. The legend on the top right panel is common for both plots. The data for this figure was obtained numerically, as described in Appendix~\ref{app:direct-numerical-trajectories}.}
    \label{fig:cumulants}
\end{figure}

It is natural to ask whether these behaviors hold in our system as well. This is a priori not obvious, since $\lambda(k)$ is only defined for $k \le 0$, so it is unclear whether its analytical series expansion at $k=0$ exists.
However, we find numerically that $\lambda(k)$ can indeed be well approximated by a perturbative power-series expansion at $0 < -k \ll 1$, as can be seen in the inset of Fig.~\ref{fig:scgf_all_maps}, in which we subtracted the linear term $\lambda_1 k$ for the sake of clarity. Furthermore, our numerical simulations show that the cumulants of $A$ grow linearly with $N$, even at high orders.
This is shown for all cumulants up to the eighth in Fig.~\ref{fig:cumulants}. 
Finally, the coefficients $\lambda_n$ of this linear growth, see Eq.~\eqref{eq:cumulants-scale-with-N}, are in excellent agreement with the derivatives $\lambda_n = \lambda^{(n)}(0)$ computed directly from the tilted operator (see Appendix~\ref{app:powerSeriesLambdak} and Table~\ref{tab:cumulants}), in accordance with Eq.~\eqref{eq:SCGF-perturbative-expansion}.

A similar situation, in which the cumulants grow linearly despite DV theory breaking down for some (or all) nonzero values of $k$ is known \cite{SmithMajumdar22, Smith24Absx}, or conjectured \cite{Smith22OU} to occur in some stochastic systems as well. An analogous situation is also known to occur in the far simpler case of large deviations of sums of i.i.d. random variables whose distributions decay slower than exponentially: In that case, the cumulants are exactly proportional to the number of summands (and the relevant theoretical framework is Cram\'{e}r's theorem rather than DV theory).

\subsection{The near right tail}

The cumulants of $P_N(A)$ are dominated by the behavior of typical fluctuations of $A$, not by the tails of the distribution. One way to see this is by expressing the cumulants using the moments (e.g., the variance $\left\langle A^{2}\right\rangle _{c}=\left\langle A^{2}\right\rangle -\left\langle A\right\rangle ^{2}$), which are dominated by the typical-fluctuations regime.
The behavior \eqref{eq:cumulants-scale-with-N} of the cumulants indicates that the DV theory does not describe just the left tail of the distribution, but also the regime of typical fluctuations (where corrections to the Gaussian approximation may be obtained from the lowest cumulants by using the Edgeworth expansion, for instance).
In fact, one may ask if the DV theory holds in the near right tail, i.e., for $A$ that is larger, but not too much larger, than $\left\langle A\right\rangle$. This will be quantified more precisely below. 
The analogous claim for sums of i.i.d. random variables whose distributions decay slower than exponentially was in fact recently shown to hold \cite{SmithCondensation25}, so long as the rate function $I(a)$ in the LDP \eqref{LDP} is written as a power series in $a - a_*$, i.e., 
$I\left(a\right)=I_{2}\left(a-a_{*}\right)^{2}+I_{3}\left(a-a_{*}\right)^{3}+\dots$ yielding
\begin{equation}
    P_{N}\left(A\right)\sim e^{-N\left[I_{2}\left(\frac{A}{N}-a_{*}\right)^{2}+I_{3}\left(\frac{A}{N}-a_{*}\right)^{3}+\dots\right]} \, . \label{eq:PNA-donsker-varadhan-extended}
\end{equation}
The series in Eq.~\eqref{eq:PNA-donsker-varadhan-extended} is then interpreted as an asymptotic series (in the limit $A - \left\langle A\right\rangle \ll N$). This interpretation is necessary, since the radius of convergence of the series $I(a)$ around $a=a_*$ is expected to be zero, given that the SCGF $\lambda(k)$ diverges for any $k>0$.

The prediction \eqref{eq:PNA-donsker-varadhan-extended}, truncating the terms beyond the 4th power in the exponential (in the left tail this truncation is unnecessary), and taking into account a normalization factor,
shows good agreement with $P_N(A)$ obtained from numerical simulations for $N=1600$, see Fig.~\ref{fig:subleading}. 
In  Appendix \ref{app:DVsubleading} we derive a subleading correction which is important at moderately large $N$ (but its effect is not visible on the scale of the figure).
These numerical simulations were performed using an importance sampling method \cite{Hartmann2025}, see Appendix \ref{app-sec:importance-sampling} for details.

\begin{figure}
    \centering
    \includegraphics[width=\linewidth]{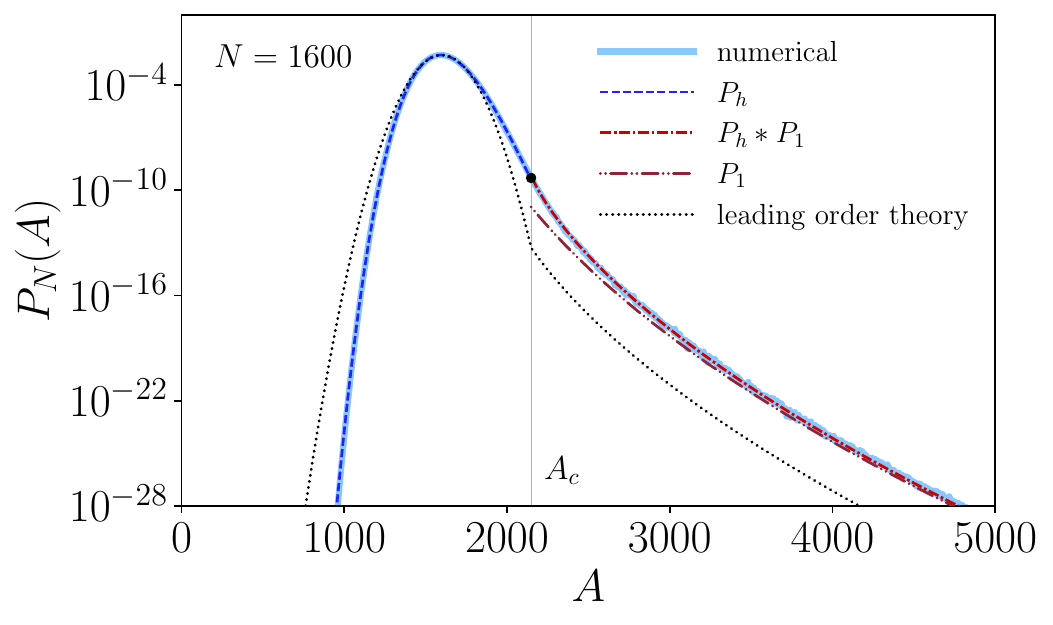}
    \caption{Distribution $P_N(A)$ for the doubling map with $N = 1600$. The solid thick line (light blue) shows results from importance-sampling simulations (see Appendix~\ref{app-sec:importance-sampling}). The small dashed line (blue), plotted up to the critical point $A_c$ (highlighted in the figure), shows the homogeneous contribution $P_h$ given by Eq.~\eqref{eq:PNA-donsker-varadhan-extended}, truncated beyond the fourth power and with a normalization factor included. See Eq.~\eqref{eq:DV-final} for the subleading prefactor. 
    The double dotted dashed line (dark brown) shows $P_1$ of Eq.~\eqref{P1A1Leading}, including the pre-exponential factor of Eq.~\eqref{stretchedExponentialDoubling}, which the distribution approaches deep in the condensed phase. The dotted line shows the leading-order prediction of Eq.~\eqref{AnomalousLDP}, illustrating the slow convergence toward the $N \to \infty$ asymptotics.
    The dot-dashed line (red) shows the convolution $P_h * P_1$ of Eq.~\eqref{eq:convolutionMainText}, evaluated numerically with the subleading-order expressions for the two distributions (see Appendix~\ref{app-sub-convolution-evaluation}).}
    \label{fig:subleading}
\end{figure}

\section{The dynamical phase transition in the right tail $\langle A \rangle < A$}
\label{sec:condensation}

Further into the right tail of the distribution, DV theory breaks down altogether, and the fluctuations behave completely differently, strongly deviating from the standard LDP scaling \eqref{LDP}.
As we show below, the correlations between the $x_i$'s give rise to a stretched exponential tail of $P_N(A)$. This tail is different to the tail of the distribution of $g(x_i)$ (which for our case is a pure exponential), and this difference is in contrast to the situation that one encounters when analyzing sums of i.i.d. random variables. In the latter case, the big jump principle implies that the tail of the distribution of the sum is, in the leading order, $N$ times larger than the tail of the distribution of each summand.

\begin{figure}
    \centering
    \includegraphics[width=\linewidth]{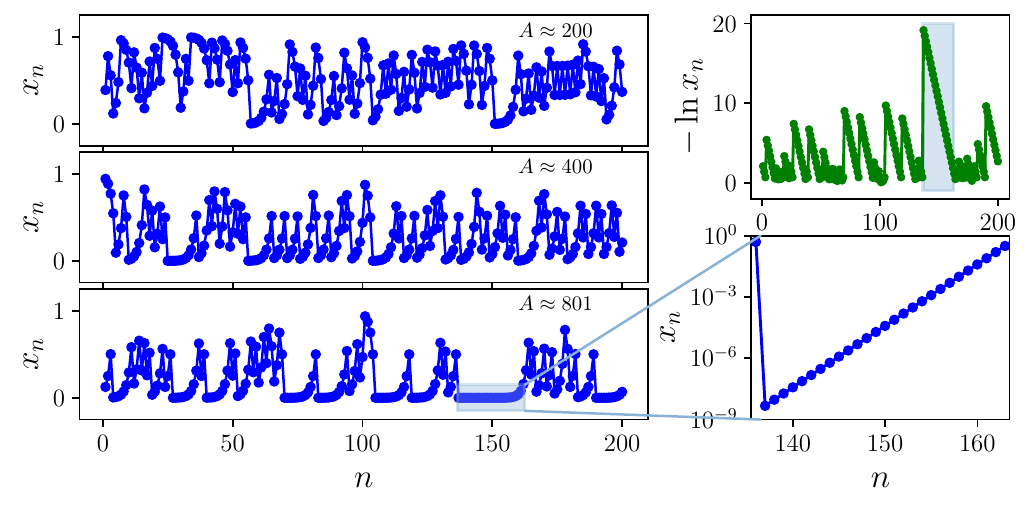}
    \caption{Trajectories $x_1,\dots,x_N$ of the doubling map for $N=200$ and three values of $A\approx 200$, $400$, and $801$ (left column, top to bottom). As $A$ grows, the trajectories spend increasingly longer stretches in the region $x\ll 1$. The right column focuses on the $A\approx 801$ trajectory: the top-right panel shows $-\ln x_n$ over the full range $0\le n\le 200$, where the shaded band (also shown in the bottom-left panel) marks a concentrated burst of large contributions to $A$, the hallmark of condensation. The bottom-right panel zooms in on this band, $135\lesssim n\lesssim 165$, showing $x_n$ on a logarithmic scale. The straight line on the log scale is the geometric growth $x_{n+m}=2^m\epsilon$ of Eq.~\eqref{eq:xmin-trajectory}: after dropping to $x_n\approx 10^{-9}$, the trajectory needs $M\approx 30$ steps to return to $x=O(1)$. This illustrates the condensed phase, in which a fraction $M/N\ll 1$ of the trajectory dominates the fluctuation.}
    \label{fig:condensation-trajectories}
\end{figure}

As we show shortly, the stretched exponential tail arises due to realizations that include a run of consecutive $x_i$'s that are all very small (i.e., much smaller than one), see Fig.~\ref{fig:condensation-trajectories} for an illustrative example. Afterwards, we show that there is a nontrivial intermediate regime in which a transition occurs at a critical value $A=A_c$. This transition separates between a ``homogeneous'' phase, in which the fluctuation is created by contributions from the entire sequence (roughly speaking), and a ``condensed'' phase, in which a relatively small part of the sequence contributes macroscopically to the fluctuation. We now begin the analysis, beginning by the far tail of $P_N(A)$.

\subsection{Stretched exponential tail -- leading order}

Since the observable function $g(x) = - \ln x$ is unbounded from above as $x$ approaches $0^+$, the value of the observable $A$ becomes larger the longer the chaotic trajectory spends near the origin. Since for the maps we are considering $x=0$ is a fixed point, if any point $x_i$ in the trajectory is small enough, $x_i \ll 1$, then there will follow a sequence of size $M \gg 1$ of points which are all small. For the doubling and tent maps, we have that $f(x) = 2 x$, as $x \to 0$. For simplicity, we focus on these two maps in this subsection. A more detailed and general treatment, which includes also subleading corrections, is given in Appendix \ref{app:condensation}.

Let us consider the minimum of the trajectory,
\begin{equation}
    x_{\min}=\min_{i\in\left\{ 1,\dots,N\right\} }x_{i} \, .
\end{equation}
We will be interested in the case $x_{\min}=\epsilon \ll 1$, as this corresponds to the mechanism described above for creating a large contribution to $A$.
Let us now calculate the number of small elements $M$ in the sequence that follow the minimum value.
Let $x_n = \epsilon$. For sufficiently small $m$, subsequent points are given by
\begin{align}
    x_{n+m} = 2^{m} \epsilon \, . \label{eq:xmin-trajectory}
\end{align}
Now $M$ is found by requiring $x_{n+M} \sim O(1)$, which yields
\begin{align}
M \simeq - \frac{\ln \epsilon}{\ln 2} + O(1) \, . \label{eq:def-of-M}
\end{align}
Let us define $A_1$  as the contribution of this subsequence, shifted by the ensemble average, that is
\begin{align}
A_{1}=-\sum_{m=0}^{M}\left(\ln x_{n+m}-\left\langle \ln x\right\rangle _{S}\right)\,.
\end{align}
Using Eqs.\,(\ref{eq:xmin-trajectory}) and (\ref{eq:def-of-M}), together with $\langle - \ln x \rangle_S = 1$ (for the doubling and tent maps), we calculate this sum as a function of $\epsilon$ and find
\begin{align}
\label{A1ofepsilon}
    A_1 \simeq \frac{(\ln \epsilon)^2}{2 \ln 2} -  \ln \epsilon \left[ \frac{1}{2} - \frac{\langle - \ln x \rangle_S}{\ln 2} \right] \, .
\end{align}

Our goal now is to calculate the distribution of $A_1$. For our purposes, it is sufficient to obtain its large-$A_1$ tail behavior, which is governed by realizations with unusually small values of $x_{\min}\ll\left\langle x_{\min}\right\rangle \sim1/N$. We therefore calculate the probability that $x_{\min} \in [0,\epsilon]$ for $\epsilon \ll 1/N$. For each $i \neq 1$, the probability that $0 < x_i < \epsilon$ and that, in addition, $x_i = x_{\min}$ is $\epsilon/2$, where the factor of $1/2$ comes from the requirement that $x_{i-1}$ be the larger of the two preimages of $x_i$ under $f$. Summing these probabilities over $i$, we find that
\begin{align}
    \mathrm{Prob}(0 < x_\mathrm{min} < \epsilon) \simeq \frac{N \epsilon}{2} \, . \label{eq:prob-of-xmin-less-epsilon}
\end{align}
Solving Eq.~\eqref{A1ofepsilon} for $\epsilon=\epsilon(A_1)$ and substituting into Eq.~\eqref{eq:prob-of-xmin-less-epsilon}, we find that the PDF of $A_1$ decays as a stretched exponential at large $A_1$,
\begin{equation}
\label{P1A1Leading}
    P_1(A_1) \sim e^{-c\sqrt{ A_1}} \, ,
\end{equation}
where (for the doubling and tent maps) $c = \sqrt{2\ln 2}$.
In Appendix \ref{app:condensation} we obtain a similar result for the logistic map (but with a different value of $c$), and also obtain the subleading correction to Eq.~\eqref{P1A1Leading}.

Importantly, at large $A_1$ the stretched-exponential decay \eqref{P1A1Leading} is far slower than the exponential decay in $N$ that governs homogeneous fluctuations [Eq.~\eqref{LDP}]. It is therefore overwhelmingly more likely to create a large fluctuation of $A$ through a single localized excursion than through a collective contribution of the entire sequence. As a result, the large-$A$ tail of $P_N(A)$ is dominated by realizations in which a localized contribution carries essentially the entire excess, $A_1 \simeq A - \langle A \rangle$, and the distribution of $A$ inherits the stretched-exponential decay
\begin{equation}
 \label{PNABigJump}
    P_N(A) \sim e^{-c\sqrt{ A - \langle A \rangle}} \, .
\end{equation}
This is qualitatively similar to the ``big-jump principle'' from the theory of distributions of sums of i.i.d. random variables or in continuous-time random walks \cite{Chistyakov64,Foss13,Denisov08,Geluk09,Clusel06, BCV10,BUV14,VBB19, WVBB19,Gradenigo13,Barkai20, MKB98, BBBJ2000, EH05, MEZ05, EMZ06, Majumdar10, CC12, ZCG, Corberi15, BVB24, HBB24, VB24, BVBB25}.
The prediction in Eq.~\eqref{PNABigJump}, together with its subleading correction (given in Appendix \ref{app:condensation}), is compared to results of numerical simulations in Fig.~\ref{fig:subleading} with very good agreement at large $A$.

In this subsection, we tacitly assumed that $M \ll N$, or equivalently that $A_1, A \ll N^2$, so that the duration of the ``big jump'' is much shorter than the duration of the entire dynamics. In the regime of very large fluctuations, $A \sim N^2$, the big jump occupies an $O(1)$ fraction of the dynamics and is no longer a localized event. Analogous regimes have been studied in other systems \cite{SmithMajumdar22, Smith24Absx}, but we leave this regime for future work.

\subsection{Condensation transition}

Let us now analyze the transition between the Gaussian regime, which describes typical fluctuations, and the big-jump regime.
In the Gaussian regime, $P_N(A)$ scales (in the leading order) as
$-\ln P_{N}\left(A\right)\sim\left(A-\left\langle A\right\rangle \right)^{2}/N$,
while in the big-jump regime,
$-\ln P_{N}\left(A\right)\sim\sqrt{A-\left\langle A\right\rangle }$.
It is therefore natural to expect a nontrivial behavior at moderately large values
$A-\left\langle A\right\rangle \sim N^{2/3}$,
where these two predictions are of the same order of magnitude.

Indeed, for $A-\left\langle A\right\rangle \sim N^{2/3}$, it is reasonable to expect that one part of the fluctuation may be created by a big jump of size $A_1 \sim N^{2/3}$, while the remainder of the fluctuation, $A_h = A  - A_1$, is created homogeneously from the rest of the dynamics. Under this assumption, the big jump involves just $M \sim \sqrt{A_1} \sim N^{1/3} \ll N$ sequence elements, so it can be viewed as a localized event, and its effect on the homogeneous contribution is negligible. In other words, the two contributions, $A_1$ and $A_h$, can be considered (approximately) statistically independent.

It is therefore straightforward to calculate the distribution of 
$ A  = A_1 + A_h$,
by using the convolution of the distributions of $A_1$ and $A_h$.
To leading-order, this calculation is relatively simple and coincides with analogous calculations which have been performed for sums of i.i.d. random variables \cite{SmithCondensation25} as well as other systems \cite{Smith22OU, SmithMajumdar22}. However, for completeness, we give here the full details of the calculation.

For simplicity, let us focus on the doubling map. Here, the right tail of the distribution of $A_1$ is a stretched exponential, as described by Eq.~\eqref{P1A1Leading}.
On the other hand, the distribution of the homogeneous contribution is described by Eq.~\eqref{eq:PNA-donsker-varadhan-extended}. In the leading order, this is simply given by the Gaussian part as described by the parabolic asymptotic behavior of $I(a)$ near its minimum, i.e.,
\begin{equation}
\label{PhAhLeading}
P_h\left(A_h\right)\sim e^{-I_{2}\left(A_h-\left\langle A\right\rangle \right)^{2}/N}\,,
\end{equation}
where the value of $I_2$ for the doubling map is given in Table \ref{tab:rate-coefficients}.

Next, we calculate the convolution of these two distributions,
\begin{equation}
    P_N(A) \simeq \int_0^{A-\left\langle A\right\rangle } d A_1\; P_1(A_1)\,P_h(A - A_1 ) \, .
    \label{eq:convolutionMainText}
\end{equation}
We now assume the scaling
$A-\left\langle A\right\rangle \sim N^{2/3}$,
so we denote
$A-\left\langle A\right\rangle =N^{2/3}y$,
and similarly, we denote
$A_{1}=N^{2/3}z$.
Using these scaling relations, and plugging Eqs.~\eqref{P1A1Leading} and \eqref{PhAhLeading} into \eqref{eq:convolutionMainText}, we obtain
\begin{align}
\label{PNAWithScalings}
    P_{N}(A)&\sim\int_{0}^{A-\left\langle A\right\rangle }dA_{1}\;e^{-c\sqrt{A_{1}}}\,e^{-I_{2}\left(A-A_{1}-\left\langle A\right\rangle \right)^{2}/N}\nonumber\\
    &\sim\int_{0}^{y}dz\;e^{-N^{1/3}\left[c\sqrt{z}+I_{2}\left(y-z\right)^{2}\right]} \, .
\end{align}
Next, we exploit the large parameter $N^{1/3} \gg 1$ in the exponential to apply the saddle-point approximation to the integral in Eq.~\eqref{PNAWithScalings}. The result takes the scaling form of an LDP with anomalous scaling exponents:
\begin{equation}
\label{AnomalousLDP}
    P_{N}(A) \sim e^{-N^{1/3} F((A-\left\langle A\right\rangle) / N^{2/3})} \, ,
\end{equation}
where the large-deviation function $F(y)$ is given by
\begin{equation}
\label{Fdef}
F\left(y\right)=\min_{z\in\left[0,y\right]}\mathcal{F}\left(y,z\right)
\end{equation}
with
\begin{equation}
\mathcal{F}\left(y,z\right)=c\sqrt{z}+I_{2}\left(y-z\right)^{2} \, .
\end{equation}

Interestingly, the large-deviation function $F(y)$  coincides, up to the scaling factors, with large-deviation functions which have been encountered in several other settings \cite{Smith22OU, SmithMajumdar22, Smith24Absx, GM19, MLMS21, MGM21, SmithCondensation25, BH26, Burov26}. Its most striking feature is a corner singularity at a critical value \cite{Smith22OU}
$y_{c}=(3/2^{5/3})\left(c/I_{2}\right)^{2/3}$.
At $0<y<y_c$, the minimizer in Eq.~\eqref{Fdef} is $z=0$, so that $F(y)$ is purely parabolic, $F\left(y\right)=I_{2}y^{2}$, describing a homogeneous phase.
At $y>y_c$, the minimizer in Eq.~\eqref{Fdef} is nonzero. It is then convenient to give $F(y)$ in a parametric form, where $F$ and $y$ are each written explicitly as functions of the minimizer $z$:
\begin{align}
    y&=z+\frac{c}{4I_{2}\sqrt{z}}\,,\\
    F&=c\sqrt{z}+\frac{c^{2}}{16I_{2}z}\,.
\end{align}

\begin{figure}
    \centering
    \includegraphics[width=\linewidth]{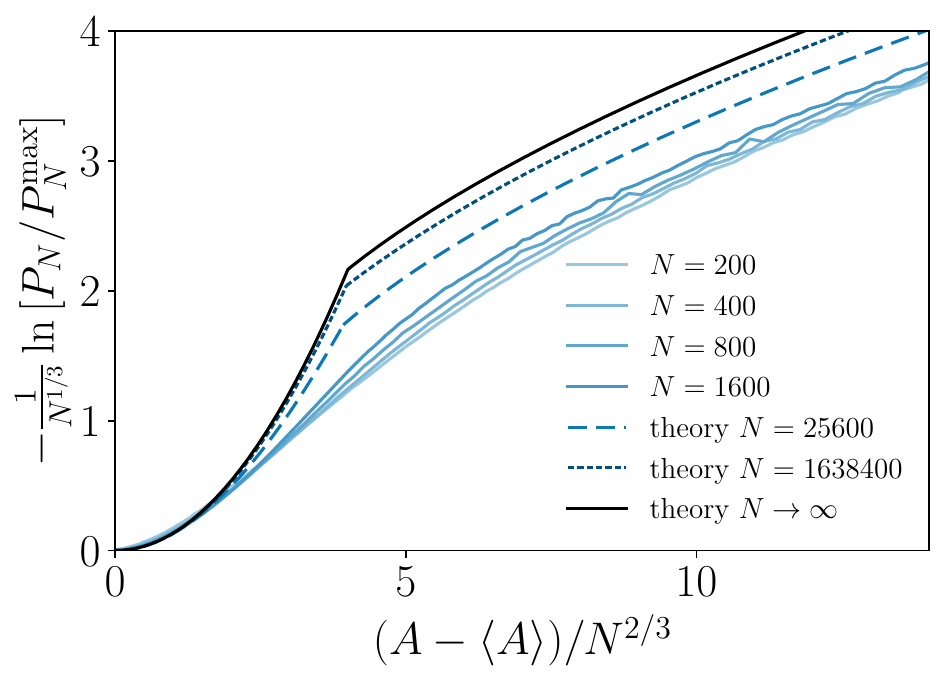}
    \caption{The right tail of the distribution $P_N(A)$, plotted as $-{N^{-1/3}}\ln\left[P_N(A)/P_N^{\mathrm{max}}\right]$ as a function of the rescaled centered variable $(A-\langle A\rangle)/N^{2/3}$, in accordance with the anomalous scaling in Eq.\,\eqref{AnomalousLDP}. Solid lines show importance-sampling simulations for $N=200$ to $1600$. For $N=25600$ and $N=1638400$, it is not feasible to perform simulations, so we instead plot the subleading-order theoretical prediction. For large $N$, the curves appear to converge to the $N\to\infty$ limit given by the large-deviation function $F(y)$ of Eq.~\eqref{Fdef} (black line), which exhibits a corner singularity.}
    \label{fig:rate_function_beyond_critical}
\end{figure}

At $y=y_c$, $F(y)$ has a corner singularity -- a jump in its first derivative, see Fig.~\ref{fig:rate_function_beyond_critical}. We interpret this singularity as a first-order dynamical phase transition (DPT), in analogy with phase transitions in equilibrium which involve singularities of the free energy \cite{Baek15, Baek17, Baek18}. This DPT describes a ``condensation'' of the fluctuation onto a localized part of the dynamics.
Finally, at $y \gg 1$, the asymptotic behavior of the large-deviation function is $F(y) \simeq c\sqrt{y}$, which provides a smooth matching with the ``big-jump'' regime. In this regime, $z \simeq y$, meaning that the fluctuation is dominated by $A_1$ (while the homogeneous contribution is subleading).

In Fig.~\ref{fig:rate_function_beyond_critical}, we compare the prediction \eqref{AnomalousLDP} with numerical simulations for moderately-large values of $N$, in the range $200$-$1600$. The trend in the numerics is indeed towards our theoretical prediction $F(y)$, but the convergence appears to be extremely slow (this is known to be the case in other systems where condensation occurs, see e.g. \cite{SmithCondensation25}). This slow convergence makes it difficult to test the theory.

We therefore developed a subleading-order theory of the transition regime $A - \left\langle A\right\rangle  \sim N^{2/3}$, see Appendix \ref{app:condensation}, and compared its predictions with numerics with excellent agreement, see Fig.~\ref{fig:subleading}. Besides enabling a comparison with numerics, this subleading order theory is also important from a practical point of view, given the slow convergence of the leading order. Most of the ingredients of our subleading-order theory are based on an analogous theory that was recently developed for studying sums of i.i.d. random variables with stretched-exponential tails \cite{SmithCondensation25}.
Predictions of our subleading-order theory are also plotted in Fig.~\ref{fig:rate_function_beyond_critical} for very large values of $N$ allowing for a quantitative assessment of the convergence rate to the leading-order result as $N$ is increased.

\section{Discussion}
\label{sec:discussion}

In summary, we studied the full distribution of dynamical observables for chaotic sequences $x_1, \dots, x_N$ of length $N \gg 1$ generated by the doubling, tent, or logistic maps. We focused on the particular case where the observable is given by minus the sum of the logarithms of the elements of the sequence. The left tail of the distribution follows the scaling \eqref{LDP}, which is commonly observed in such systems. We calculate the associated rate function by applying the extension of DV theory to discrete chaotic dynamics \cite{Smith2022}.

However, the right tail behaves very differently: It departs from the scaling \eqref{LDP}. This is the consequence of the fact that the observable is unbounded from above \cite{NT20, Pontes26}.
We find that the right tail is described instead by the anomalous scaling \eqref{AnomalousLDP}, and we calculated the large-deviation function $F(y)$ analytically. Interestingly, the large-deviation function has a singularity which we interpret as a first-order dynamical phase transition (DPT). The DPT separates a homogeneous phase, in which the fluctuation arises due to contributions that are of the same order of magnitude from all elements of the sequence, and a ``condensed'' phase in which a small fraction of the $x_i$'s give a macroscopic contribution. We also find that the cumulants of the distribution grow linearly with $N$ (at $N \gg 1$).
Strikingly, these same phenomena of anomalous scaling, with the same large-deviations function up to scaling factors, as well as a linear growth of the cumulants, have been observed in several other systems \cite{Smith22OU, SmithMajumdar22, Smith24Absx, GM19, MLMS21, MGM21, SmithCondensation25, BH26, Burov26}, suggesting that the behavior of fluctuations in these systems is universal.

We supported our theoretical predictions through  extensive numerical simulations, using importance-sampling techniques. We found that, in the right tail, the convergence of our asymptotic results is very slow in the limit $N \to \infty$. We therefore developed a subleading-order theory, in analogy with the one recently developed for studying sums of i.i.d. random variables with stretched-exponential tails \cite{SmithCondensation25}, and this enabled us to verify our results at moderately-large $N$. Nevertheless, given that the analytical methods we used in this work are not mathematically rigorous, it could be useful to re-derive our results in a more rigorous fashion.

This research opens up several avenues for future research. One could consider unbounded observables in chaotic dynamics in a higher spatial dimension and/or in continuous time. It would also be interesting to study observables whose distribution tails decay as a power law, rather than as a stretched exponential as in the present paper.
Finally, it could be useful to develop subleading-order theories of the condensation transition analogous to the ones derived here and in \cite{SmithCondensation25} for other systems in which the condensation transition is expected to occur.

\bigskip

\section*{Acknowledgments}

We acknowledge support from the Israel Science Foundation (ISF) through Grant No. 2651/23,  and from the Binational Science Foundation (BSF) through Grant No. 2024162. NRS acknowledges support from the Golda Meir fellowship.

\bigskip

\appendix

\section{Numerical calculation of the SCGF $\lambda(k)$ \label{app:numerical-SCGF-power-method}}

Our goal in this Appendix is to explain the numerical methods used to calculate the SCGF $\lambda(k)$, defined directly from Eq.~(\ref{eq:SCGF}), and as the largest (right) eigenvalue of the tilted operator in Eq.~(\ref{eq:tilted-operator-definition}), which we repeat here for convenience 
\begin{equation} 
L_{k}\psi_{R}(x)=\sum_{z\in f^{-1}(x)}\frac{e^{kg(z)}\psi_{R}(z)}{\left|f'(z)\right|} \, . 
\end{equation} 
We use the power method, which is a standard iterative algorithm for finding the largest eigenvalue and eigenvector of a linear operator \cite{Trefethen1997, Saad2011, Golub2013}. The idea is straightforward: we apply the operator $L_k$ repeatedly to an initial function $\varphi^{(0)}(x)$, which we take to be the invariant measure $p_s(x)$. The component of $\varphi^{(0)}$ along the leading eigenfunction grows by a factor of $e^{\lambda(k)}$ at each iteration, while all other components grow more slowly since their associated eigenvalues are smaller in magnitude. After sufficiently many iterations, the output is therefore dominated by the leading eigenfunction, and the ratio of successive norms converges to $e^{\lambda(k)}$. To prevent numerical overflow, the function is renormalised after each iteration, and the eigenvalue is extracted from the norm before renormalisation. Explicitly, at the $n$-th iteration: 
\begin{enumerate} 
\item Start with $\varphi^{(n)}(x)$, normalized so that $\int_0^1 \varphi^{(n)}(x)\,dx = 1$. 

\item Apply the tilted operator: $\tilde{\varphi}^{(n+1)}(x) = L_{k}\,\varphi^{(n)}(x)$. 

\item Extract the eigenvalue estimate from the norm: 
    \begin{equation} 
    e^{\tilde{\lambda}^{(n+1)}(k)} \equiv \int_0^1 \tilde{\varphi}^{(n+1)}(x)\,dx \, . 
    \end{equation} 
\item Renormalise for the next iteration: $\varphi^{(n+1)}(x) = \tilde{\varphi}^{(n+1)}(x)\,e^{-\tilde{\lambda}^{(n+1)}(k)}$. \end{enumerate} Steps 2--4 are repeated until $|\tilde{\lambda}^{(n+1)}(k) - \tilde{\lambda}^{(n)}(k)| < \varepsilon$ for some prescribed tolerance $\varepsilon$, at which point $\tilde{\lambda}^{(n+1)}(k)$ is taken as the estimate of $\lambda(k)$.

\smallskip

\subsection{The discretization of the operator}
For $g(x) = -\ln x$ and for the maps under consideration, the operator $L_k$ takes the explicit form
\begin{align}
    L_k \varphi(x) = \frac{z_1^{-k}\,\varphi(z_1)}{|f'(z_1)|}
    + \frac{z_2^{-k}\,\varphi(z_2)}{|f'(z_2)|} \, ,
\end{align}
where $z_1$ and $z_2$ are the two pre-images of $x$ under the map $f$.
We discretize the domain $x \in [0,1]$ using a composite grid: for $0 < x < 0.2$ we use a logarithmically spaced grid starting from a small value $x_{m}$, and for $0.2 < x < 1$ we use a uniformly spaced grid. The motivation for this choice will become clear in the next subsection. Since the pre-image points $z_{1,2}$ generically do not coincide with grid nodes, their function values are estimated using a cubic spline interpolation algorithm that supports non-uniform grids. For pre-image values that fall outside the grid, for instance when $z_{1,2} < x_{m}$, the same spline is used in extrapolation mode.

\medskip

\subsection{The derivatives of $\lambda(k)$ via a power series expansion}

\label{app:powerSeriesLambdak}

To calculate the rate function and the cumulants of the observable, we need the derivatives of $\lambda(k)$ at a chosen point $k_0$. While low-order derivatives are straightforward to obtain, high-order ones (up to eighth order in this work) require a level of numerical precision that is difficult to achieve by direct finite differencing. We resolve this by performing a Taylor expansion of the entire eigenvalue problem around $k_0$, so that all derivatives are extracted simultaneously from a single run of the power iteration. We expand the SCGF as
\begin{align}
    \lambda(k_0 + \delta k) = \lambda(k_0) + \sum_{n=1}^{\infty} \frac{\lambda^{(n)}(k_0)}{n!}\,\delta k^n \, ,
\end{align}
where $\delta k$ is a formal expansion parameter whose numerical value is never specified. The weight factor appearing in the tilted operator \eqref{eq:tilted-operator-definition} is expanded as
\begin{align}
    e^{-(k_0+\delta k)\ln z} = z^{-k_0}\sum_{n=0}^{\infty}\frac{(-\ln z)^n}{n!}\,\delta k^n \, ,
\end{align}
which is why the log-spaced grid near the origin is essential: the coefficients $(-\ln z)^n$ grow rapidly as $z\to 0$, and the log-spacing ensures they have proper numerical resolution. 

\begin{table}[h]
\centering
\renewcommand{\arraystretch}{1.6}
\begin{tabular}{lcccc}
\toprule
Map & $f(x)$ & $z_1$ & $z_2$ & $1/|f'(z_{1,2})|$ \\
\midrule
Doubling & $f=2x\bmod 1$ & $x/2$ & $(1+x)/2$ & $1/2$ \\[2pt]
Tent & $f=1-|2x-1|$       & $x/2$ & $1-x/2$   & $1/2$ \\[2pt]
Logistic & $f=4x(1-x)$   & $\frac{1-\sqrt{1-x}}{2}$ &
                          $\frac{1+\sqrt{1-x}}{2}$ &
                          $\frac{1}{4\sqrt{1-x}}$ \\
\bottomrule
\end{tabular}
\caption{Pre-images and Jacobian factors for all considered maps.}
\label{tab:preimages}
\end{table}

The eigenfunction is likewise expanded as
\begin{align}
    \varphi(x)
    = \varphi_0(x) + \varphi_1(x)\,\delta k
    + \varphi_2(x)\,\delta k^2 + \cdots
\end{align}
where $\varphi_n(x)$ is the $n$-th Taylor coefficient of the eigenfunction, to be determined iteratively. Substituting these expansions into the operator equation for the doubling map (see Table \ref{tab:preimages}) and collecting powers of $\delta k$, we write the update rule for the first three orders as
\begin{widetext}
\begin{align}
    \tilde{\varphi}_0(x) &=
    \frac{z_1^{-k_0}\,\varphi_0(z_1)
    + z_2^{-k_0}\,\varphi_0(z_2)}{2} \, , \\
    \tilde{\varphi}_1(x) &=
    \frac{z_1^{-k_0}\bigl[\varphi_1(z_1)
    + (-\ln z_1)\,\varphi_0(z_1)\bigr]
    + z_2^{-k_0}\bigl[\varphi_1(z_2)
    + (-\ln z_2)\,\varphi_0(z_2)\bigr]}{2} \, , \\
    \tilde{\varphi}_2(x) &=
    \frac{z_1^{-k_0}\Bigl[\varphi_2(z_1)
    + (-\ln z_1)\,\varphi_1(z_1)
    + \tfrac{(-\ln z_1)^2}{2}\,\varphi_0(z_1)\Bigr]
    + z_2^{-k_0}\Bigl[\varphi_2(z_2)
    + (-\ln z_2)\,\varphi_1(z_2)
    + \tfrac{(-\ln z_2)^2}{2}\,\varphi_0(z_2)\Bigr]}{2} \, .
\end{align}
\end{widetext}
A clear pattern emerges where the update for $\tilde{\varphi}_n$ is strictly a function of equal- and lower-order coefficient functions $\varphi_0, \ldots, \varphi_{n-1}$ through the series. After the operator is applied, we compute the norm of the updated  eigenfunction
\begin{align}
    Z(k_0 + \delta k) = \int_0^1 \tilde{\varphi}(x)\,dx
    = e^{\lambda(k_0+\delta k)} \, ,
\end{align}
where the last equality holds when the approximation has converged, since integrating the eigenvalue equation $L_k\varphi = e^{\lambda(k)}\varphi$ over $x$ with the normalisation $\int\varphi(x)\,dx = 1$ gives exactly $Z = e^{\lambda}$. The SCGF and all its derivatives at $k_0$ are therefore encoded in $Z$.
Expanding $Z$ as a power series in $\delta k$, its Taylor coefficients are simply the norm integrals of the individual eigenfunction coefficients,
\begin{align}
    Z(k_0 + \delta k) = \sum_{n=0}^{\infty} Z_n\,\delta k^n \, ,
    \label{ZnFromPhi}
    \qquad Z_n = \int_0^1 \tilde{\varphi}_n(x)\,dx \, ,
\end{align}
from which the SCGF derivatives are recovered as $\lambda^{(n)}(k_0) = n!\,\ell_n$, where the $\ell_n$ are the Taylor coefficients of $\ln Z$. 
As a concrete example, for the second order of the doubling map, one begins from calculating $\tilde{\varphi}_0$, $\tilde{\varphi}_1$, $\tilde{\varphi}_2$, the updated eigenfunction coefficients from the previous step. From them, $Z_0$, $Z_1$ and $Z_2$ are calculated using Eq.~\eqref{ZnFromPhi}.
The log-Taylor recurrence then gives
\begin{align}
    \ell_0 &= \ln Z_0 \, , \\
    \ell_1 &= \frac{Z_1}{Z_0} \, , \\
    \ell_2 &= \frac{1}{2Z_0}\left(2Z_2 - \ell_1 Z_1\right)
             = \frac{Z_2}{Z_0} - \frac{Z_1^2}{2Z_0^2} \, ,
\end{align}
from which the eigenvalue estimates at this iteration are
\begin{align}
    \tilde{\lambda}(k_0) &= \ell_0 = \ln Z_0 \, , \\
    \tilde{\lambda}'(k_0) &= 1!\,\ell_1 = \frac{Z_1}{Z_0} \, , \\
    \tilde{\lambda}''(k_0) &= 2!\,\ell_2 = \frac{2Z_2}{Z_0} - \frac{Z_1^2}{Z_0^2} \, .
\end{align}
This procedure extends straightforwardly to any general order $M$. At each
iteration one computes $Z_0, \ldots, Z_M$ and applies the recurrence to obtain $\ell_0, \ldots, \ell_M$, recovering all derivatives $\tilde{\lambda}^{(n)}(k_0) = n!\,\ell_n$ for $n = 0, \ldots, M$ simultaneously. In this work we carry the expansion to $M = 8$.
Convergence is checked through the highest-order estimate, since
$\tilde{\lambda}^{(M)}$ is the last to converge. The iteration is stopped when
\begin{align}
    \left|\tilde{\lambda}^{(M),(\mathrm{new})}(k_0)
    - \tilde{\lambda}^{(M),(\mathrm{old})}(k_0)\right| < \varepsilon \, ,
\end{align}
for a prescribed tolerance $\varepsilon = 10^{-5}$, at which point all lower-order derivatives are guaranteed to have converged as well. After the convergence check, the eigenfunction coefficients are updated by normalising $\tilde{\varphi}_n$ by $Z\left(k_0 + \delta k \right)$ as a formal power series. For $M=2$, the inverse coefficients $d_n$ of $1/Z(k_0 + \delta k)$ are
\begin{align}
    d_0 &= \frac{1}{Z_0} \, , \\
    d_1 &= -\frac{Z_1}{Z_0^2} \, , \\
    d_2 &= -\frac{1}{Z_0}\left( d_1 Z_1 + d_0 Z_2 \right)
          = \frac{Z_1^2}{Z_0^3} - \frac{Z_2}{Z_0^2} \, ,
\end{align}
and the updated eigenfunction coefficients are
\begin{align}
    \varphi_0(x) &\leftarrow d_0\,\tilde{\varphi}_0(x) \, , \\
    \varphi_1(x) &\leftarrow d_0\,\tilde{\varphi}_1(x)
                            + d_1\,\tilde{\varphi}_0(x) \, , \\
    \varphi_2(x) &\leftarrow d_0\,\tilde{\varphi}_2(x)
                            + d_1\,\tilde{\varphi}_1(x)
                            + d_2\,\tilde{\varphi}_0(x) \, .
\end{align}

It is important to highlight that no specific value of $\delta k$ is ever used as the entire procedure operates on the coefficients of the expansion and $\delta k$ is a formal parameter throughout.
We carried out this calculation explicitly around $k_0=0$, yielding the cumulants $\lambda_n$, for the doubling, tent, and logistic maps up to the 8th order, and the results are given in Table \ref{tab:cumulants}.

\begin{table}[h]
\centering
\begin{tabular}{lccc}
\hline\hline
Cumulant & Doubling & Tent & Logistic \\
\hline
$\lambda_1$ & $1$                     & $1$                     & $\ln 4$                     \\
$\lambda_2$ & $3.705486$                     & $2.715004$                     & $\pi^2 \simeq 9.869604$                     \\
$\lambda_3$ & $5.640734 \times 10^{1}$       & $4.922782 \times 10^{1}$       & $3.568410 \times 10^{2}$       \\
$\lambda_4$ & $2.115899 \times 10^{3}$       & $1.940097 \times 10^{3}$       & $2.801931 \times 10^{4}$       \\
$\lambda_5$ & $1.380435 \times 10^{5}$       & $1.291812 \times 10^{5}$       & $3.714303 \times 10^{6}$       \\
$\lambda_6$ & $1.351917 \times 10^{7}$       & $1.279757 \times 10^{7}$       & $7.323329 \times 10^{8}$       \\
$\lambda_7$ & $1.834087 \times 10^{9}$       & $1.749541 \times 10^{9}$       & $1.990402 \times 10^{11}$      \\
$\lambda_8$ & $3.274588 \times 10^{11}$      & $3.141206 \times 10^{11}$      & $7.077667 \times 10^{13}$      \\
\hline\hline
\end{tabular}
\caption{SCGF series expansion coefficients $\lambda_n = \lambda^{(n)}(0)$ for the doubling, tent, and logistic maps. $\lambda_1 = a_*$ is calculated analytically, and for the logistic map, $\lambda_2$ is too, see Appendix\,\ref{app:exact-green-kubo-logistic}. The rest of the $\lambda_n$'s are computed by using the power-series expansion described in Appendix \ref{app:powerSeriesLambdak}.}
\label{tab:cumulants}
\end{table}

\subsection{The particular case of the logistic map, exact second cumulant via Green--Kubo \label{app:exact-green-kubo-logistic}}
The logistic map is conjugate to a piecewise-linear map through the change of variables
\begin{equation}
  x=\sin^2\frac{\theta}{2} \, ,\qquad \theta\in[0,\pi] \, ,
  \label{eq:GK-chebyshev}
\end{equation}
where $\theta$ is referred to as the Chebyshev coordinate. Writing $x_n=\sin^2(\theta_n/2)$, the iteration $x_{n+1}=f(x_n)$ gives $x_{n+1}=4\sin^2(\theta_n/2)\cos^2(\theta_n/2)=\sin^2\theta_n$, and imposing that $x_{n+1}=\sin^2(\theta_{n+1}/2)$ with $\theta_{n+1}\in[0,\pi]$ fixes the dynamics to be the tent map on $[0,\pi]$,
\begin{equation}
  \theta_{n+1}=
  \begin{cases}
    2\theta_n \, , & \theta_n\le\pi/2 \, ,\\[1mm]
    2\pi-2\theta_n \, , & \theta_n>\pi/2 \, .
  \end{cases}
  \label{eq:GK-theta-tent}
\end{equation}
In this coordinate the invariant density is uniform, $p_s(x)\,dx=dx/[\pi\sqrt{x(1-x)}\,]=d\theta/\pi$, and the Jacobian factor $1/(4\sqrt{1-x})$ of the tilted operator in Eq.~\eqref{eq:tilted-operator-definition}, which diverges as $x$ approaches $1$, is replaced by the constant factor $1/2$ of the tent map. These regularizations are the reason all numerical computations involving the logistic map in this work are performed in the Chebyshev coordinates.

However, the advantage of the Chebyshev coordinates is not only numerical. As we will show now, the observable admits an explicit Fourier expansion in $\theta$, and the tent map in Eq.~\eqref{eq:GK-theta-tent} acts on the Fourier modes without mixing them. We write Eq.~\eqref{eq:GK-chebyshev} in the form $x=(1-\cos\theta)/2$, which factorizes over complex exponentials as
\begin{equation}
  x=\frac{2-e^{i\theta}-e^{-i\theta}}{4} = \frac{\big(1-e^{i\theta}\big)\big(1-e^{-i\theta}\big)}{4} \, .  \label{eq:GK-factorized}
\end{equation}
Taking minus the logarithm of Eq.~\eqref{eq:GK-factorized} and expanding each factor with the series $-\ln(1-z)=\sum_{n\ge1}z^{n}/n$ at $z=e^{\pm i\theta}$ gives
\begin{align}
  g(x) =-\ln x  &=\ln 4+\sum_{n=1}^{\infty}\frac{e^{in\theta}+e^{-in\theta}}{n}  \nonumber \\
  &=\ln 4+2\sum_{n=1}^{\infty}\frac{\cos(n\theta)}{n} \, .  \label{eq:GK-g-fourier}
\end{align}
Stationary averages reduce to integrals over the uniform density, which is constant in Chebyshev coordinates, for which the cosine modes satisfy $\left\langle \cos(n\theta) \right\rangle_S=0$ and the product obeys the orthogonality relation $\left\langle \cos(n\theta)\cos(m\theta) \right\rangle_S=\delta_{nm}/2$ for $n,m\ge1$. The mean of the observable follows immediately from Eq.~\eqref{eq:GK-g-fourier}, $\left\langle g(x) \right\rangle_S=\ln 4$, and the variance is
\begin{equation}
  \left\langle g(x)^2 \right\rangle_S - \left\langle g(x) \right\rangle_S^2
  = 4\sum_{n=1}^{\infty}\frac{1}{2n^{2}}
  = \frac{\pi^{2}}{3} \, .
  \label{eq:GK-variance-single}
\end{equation}

We now evaluate the Green-Kubo formula \eqref{eq:variance-green-kubo}. Since the dynamics is stationary, the summand is invariant under $\delta\to-\delta$, so that
\begin{align}
  \lambda_2 =& \left\langle g(x)^2 \right\rangle_S - \left\langle g(x) \right\rangle_S^2
  + \nonumber \\
  & + 2\sum_{\delta=1}^{\infty}
  \left[ \left\langle g(x_{0})g(x_{\delta}) \right\rangle_S
  - \left\langle g(x) \right\rangle_S^2 \right] \, ,
  \label{eq:GK-one-sided}
\end{align}
where $x_{\delta}$ is obtained from $x_0$ by $\delta$ applications of the map and the term outside the sum is given by Eq.~\eqref{eq:GK-variance-single}. The correlation functions are where the Fourier expansion simplifies the calculation immensely, because the tent map Eq.~\eqref{eq:GK-theta-tent} acts on the cosine modes without mixing them. Explicitly, on the increasing branch $\cos(m\theta_{n+1})=\cos(2m\theta_n)$ directly, and on the folded branch $\cos\big(m(2\pi-2\theta_n)\big)=\cos(2m\theta_n)$. Each application of the map therefore doubles the argument of every mode, so after $\delta$ steps,
\begin{equation}
  \cos(m\theta_{n+\delta}) = \cos\big(2^{\delta} m\, \theta_n\big) \, .
  \label{eq:GK-mode-doubling}
\end{equation}
Writing $g(x)$ as in Eq.~\eqref{eq:GK-g-fourier} and plugging Eq.~\eqref{eq:GK-mode-doubling} into the correlation function, the orthogonality relation retains only the pairs of modes with $n=2^{\delta}m$, and
\begin{align}
  &\left\langle g(x_{0})g(x_{\delta}) \right\rangle_S - \left\langle g(x) \right\rangle_S^2 = \nonumber\\
  &\qquad = 4 \sum_{n,m=1}^{\infty} \frac{\left\langle \cos(n\theta_0)\cos(2^{\delta}m\,\theta_0) \right\rangle_S}{nm} \nonumber\\
  &\qquad = 4\sum_{m=1}^{\infty}\frac{1}{2\,(2^{\delta}m)\,m}
  = \frac{2^{-\delta}\pi^{2}}{3} \, .
  \label{eq:GK-correlation}
\end{align}
The correlations thus decay geometrically with the separation $\delta$. Replacing Eqs.\,\eqref{eq:GK-variance-single} and \eqref{eq:GK-correlation} into Eq.~\eqref{eq:GK-one-sided} and summing the geometric series yields
\begin{equation}
  \lambda_2 = \frac{\pi^{2}}{3}\left(1+2\sum_{\delta=1}^{\infty}2^{-\delta}\right) = \pi^{2} \, .
  \label{eq:GK-lambda2-result}
\end{equation}
This value is in agreement with our numerical findings of Appendix \ref{app:numerical-SCGF-power-method}, where we obtain a series expansion of the SCGF.

\subsection{The particular case of the logistic map, exact solution for integer negative $k$ \label{app:negative-integer-k}}

Our goal now is to find the eigenvalues for the specific cases where $k=-n$ with $n$ being a positive integer. The eigenvalue problem, Eq.~\eqref{eq:tilted-operator-definition}, for the logistic map ($f(x)=4x(1-x)$), for these values of $k$, admits a closed-form solution. We start by introducing the ansatz $\phi_k(x)$, defined implicitly from the right eigenfunction as
\begin{equation}
  \psi_R(x) = x^{k}\, p_s(x)\, \phi_k(x),
\end{equation}
where $p_s(x) = 1/\left(\pi \sqrt{x(1-x)}\right)$ is the invariant measure. The two preimages $x_\pm=\left(1\pm\sqrt{1-x}\right)/2$ satisfy $x_++x_-=1$ and $x_+x_-=x/4$, so that $p_s(x_+)=p_s(x_-)=2/\left(\pi\sqrt{x}\right)$, the Jacobian and invariant-measure factors cancel and  Eq.~\eqref{eq:tilted-operator-definition} is reduced to the far simpler equation
\begin{equation}
  \frac{x^{-k}}{2}\big[\phi_k(x_+)+\phi_k(x_-)\big]
  = e^{\lambda(k)}\,\phi_k(x).
  \label{eq:conj-eig}
\end{equation}
At $k=0$ this returns the Frobenius--Perron operator, and the solution is $\phi_0\equiv1$ and $\lambda(0)=0$.
 
Setting $k=-n$, the prefactor $x^{-k}=x^{n}$ is polynomial, and the operator in Eq.~\eqref{eq:conj-eig} maps polynomials to polynomials. As we show below, it will be useful to calculate the left-hand side of Eq.~\eqref{eq:conj-eig} for the polynomial $\phi_k(x)$. To do so, let us first calculate the $m$th power of each of the preimages $x_\pm$ by using the binomial expansion
\begin{align}
    x_\pm^m = \frac{1}{2^{m}} \sum_{m'=0}^m (\pm 1)^{m'}\,
    \frac{m!\, \left( 1-x \right)^{m'/2}}{m'!(m-m')!} \, .
\end{align}
Now, since $\phi_k$ enters in Eq.~\eqref{eq:conj-eig} only through
$\phi_k(x_+)+\phi_k(x_-)$, the odd powers of $\sqrt{1-x}$ cancel out and we obtain that the sum
\begin{align}
\label{xpmPlusxmm}
  x_+^{m}+x_-^{m} = \frac{1}{2^{m-1}} \sum_{m'=0}^{\left\lfloor m/2 \right\rfloor}
  \frac{m!\, \left( 1-x \right)^{m'}}{(2m')!(m-2m')!} \, ,
\end{align}
is a polynomial of degree $\lfloor m/2\rfloor$ in $1-x$, and hence in $x$.

Let us determine the lowest- and highest-nonzero rank terms of the polynomial \eqref{xpmPlusxmm}.
Its lowest rank is $x^0$, only the $m'=0$ term survives at $x=0$, giving
$x_+^m+x_-^m = 1 + \mathcal{O}(x)$ for every $m\ge1$ (equivalently $x_+\to1$,
$x_-\to0$ as $x\to0$), while its highest rank is $x^{\lfloor m/2\rfloor}$.
Writing the eigenfunction as 
a polynomial
\begin{equation}
\label{polynomialAnsatz}
    \phi_{-n}(x)=\sum_{m=m_{\min}}^{m_{\max}}c_{m}x^{m} \, ,
\end{equation}
where $m_{\min}$ and $m_{\max}$ are the lowest- and highest-nonzero rank terms, so that $c_{m_{\min},m_{\max}} \ne 0$,
the left-hand side
of Eq.~(\ref{eq:conj-eig}) is 
\begin{equation}
\label{lhsSumOfM}
    \tfrac{x^n}{2} \sum_{m=m_{\min}}^{m_{\max}} c_m (x_+^m+x_-^m)
\end{equation}
the constant term of every term in the sum \eqref{lhsSumOfM} contributes at order $x^n$, so its lowest nonzero-rank term is at least  $x^{n}$. The highest nonzero-rank term of the polynomial \eqref{lhsSumOfM} is $x^{\,n+\lfloor m_{\max}/2\rfloor}$,
set by the leading monomial $x^{m_{\max}}$.
The right-hand side of Eq.~(\ref{eq:conj-eig}),
$e^{\lambda(-n)}\phi_{-n}(x)$, spans the ranks $x^{m_{\min}}$ through
$x^{m_{\max}}$, so the two sides can agree only if their lowest and highest
ranks coincide. Matching the lowest rank forces $m_{\min}=n$, and matching the
highest requires $m_{\max}=n+\lfloor m_{\max}/2\rfloor$,
and it turns out that the required solution is with $m_{\max}=2n-1$.
The eigenfunction $\phi_{-n}(x)$ is therefore a polynomial supported on the ranks $x^{n}$
through $x^{2n-1}$. 

Now that $m_{\min}$ and $m_{\max}$ have been determined, one simply plugs the ansatz \eqref{polynomialAnsatz} into Eq.~\eqref{eq:conj-eig}. This yields an eigenvalue problem for the eigenvector of coefficients $\left(c_{m_{\min}},c_{m_{\min}+1},\dots,c_{m_{\max}}\right)$, i.e., the problem boils down to that of finding the eigenvector of a matrix of rank
$m_{\max}-m_{\min}+1=n$ with largest real part.
For $k=-1,-2,-3,-4$ (and so forth), this procedure yields the values $\lambda(k)$ given in Eq.~\eqref{lambdam1m2} of the main text and directly below it. The corresponding eigenfunctions (not normalized) are
\begin{align}
  \phi_{-1}(x) &= x \, ,  \\
  \phi_{-2}(x) &= x^{2} - \frac{7-\sqrt{17}}{8}\,x^{3} \, ,  \\
  \phi_{-3}(x) &\simeq x^{3} - 0.285233\,x^{4} - 0.270420\,x^{5} \, ,  \\[1mm]
  \phi_{-4}(x) &\simeq x^{4} + 0.033202\,x^{5} - 0.766152\,x^{6} + 0.055439\,x^{7} \, .
  \label{eq:eigenfunctions-negative-k}
\end{align}

\subsection{Computation of the cumulants from direct numerical simulations \label{app:direct-numerical-trajectories}}

To validate the results obtained from the power iteration method, we compare the cumulants of $A$ against those obtained from direct trajectory-based simulations. As described in Appendix \ref{app-sec:importance-sampling}, trajectories are generated using the backward-branch representation, which avoids the exponential amplification of floating-point errors inherent to forward iteration and renders the trajectory space explicitly stochastic.

\begin{figure}
    \centering
    \includegraphics[width=\linewidth]{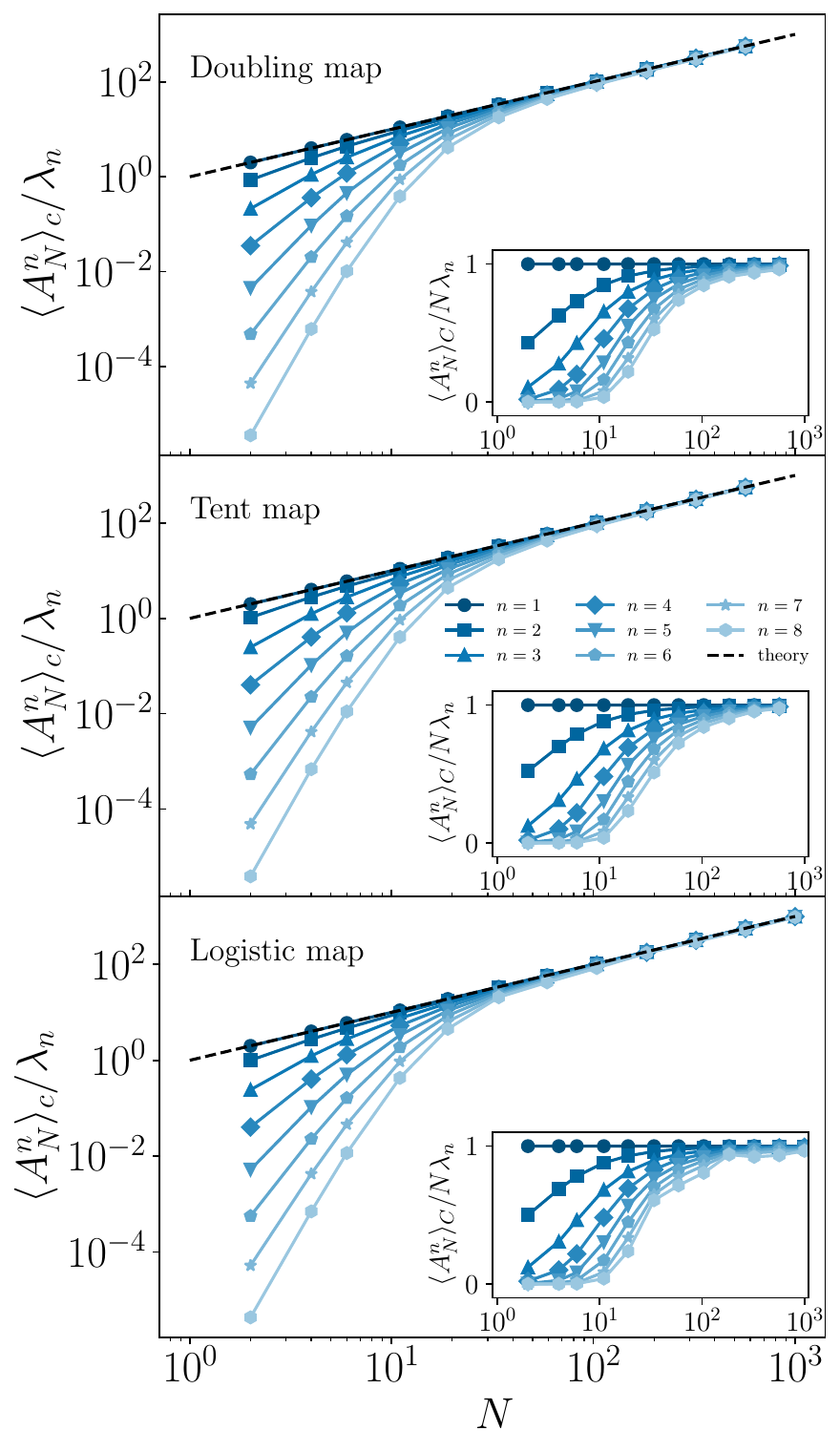}
    \caption{Numerically-computed cumulants $\langle A^n_N \rangle_c$ of the observable $A$, scaled by the coefficients $\lambda_n$, as functions of the trajectory length $N$, for the doubling (top), tent (middle) and logistic (bottom) maps. Symbols show the results of direct numerical simulations, obtained as described in Appendix~\ref{app:direct-numerical-trajectories}, for orders $n = 1, \ldots, 8$. The dashed lines represent the linear-growth prediction $\langle A^n_N \rangle_c \simeq \lambda_n N$ of Eq.~\eqref{eq:cumulants-scale-with-N}, with the values of $\lambda_n$ computed from the tilted operator via the power-series method of Appendix~\ref{app:powerSeriesLambdak} and given in Table~\ref{tab:cumulants}. The insets show the same data divided by $N\lambda_n$, demonstrating on a linear scale the convergence toward the predicted linear growth at large $N$. The legend is common to all three panels.}
    
    \label{fig:all-moments-appendix}
\end{figure}

Along each trajectory of length $N$, the observable $A$ is computed, and the raw moments $\langle A^n \rangle$ for $n = 1, \ldots, 8$ are estimated by averaging over $N_S$ independent trajectories, accumulated in extended precision to reduce cancellation errors at high order. The cumulants are then extracted from the moments via the standard moment-cumulant relations (see, e.g., \cite{MomentsCumulantsWiki}). This procedure is repeated for a logarithmically spaced set of trajectory lengths $N$, allowing the linear growth $\langle A^n \rangle_c \simeq \lambda^{(n)} N$ to be verified directly and the growth rates to be compared against the derivatives $\lambda^{(n)}(0)$ obtained from the power iteration. This is shown in Figs.~\ref{fig:cumulants} and \ref{fig:all-moments-appendix}.

\section{Importance sampling for the tails of $P_N(A)$ \label{app-sec:importance-sampling}}

Since the probability of very large or very small values of $A$ cannot be found effectively through naive sampling, we employ an importance sampling strategy to obtain a complete expression for the PDF $P_N(A)$. First, we will introduce the backward-branch representation of chaotic trajectories \cite{Smith2022, DS26}, which provides both numerical stability and a natural stochastic structure that makes Monte Carlo sampling well-defined, and then will describe the Metropolis scheme used to bias the sampling toward the tails of the distribution.

\subsection{The backward-branch representation}

Forward iteration of chaotic maps is numerically unstable, as the positive Lyapunov exponent ensures that any floating-point rounding error is amplified exponentially at each step~\cite{Smith2022, DS26}. We shall therefore build our trajectories backwards by repeated application of the inverse branches of $f(x)$. 
First, we sample the final point $x_N$ from the IM. Next, we generate the chaotic sequence in reverse order by sampling $x_{n-1}$ given $x_n$ as follows.
At each step $n$, a bit $b_n \in \{0,1\}$ selects one of the two pre-image branches, and the full trajectory $x_1, x_2, \ldots, x_N$ is uniquely determined by the initial condition $x_1$ and the bit sequence $\mathbf{b}$. Since each inverse branch is a contraction by a factor of $1/|f'|$, floating-point errors are suppressed at every step rather than amplified, and the trajectory remains numerically valid for arbitrarily many steps. Given that all maps under consideration have two branches of equal probability, naive uniform sampling of the bits produces trajectories whose elements are each distributed according to the invariant measure $p_s(x)$.

Beyond numerical stability, the backward representation offers a deeper advantage: it renders the trajectory space stochastic. Sampling $\mathbf{b}$ uniformly over $\{0,1\}^N$ is equivalent to sampling trajectories from $p_s(x)$, since $p_s$ is invariant under $f$. This reframing maps the problem of sampling chaotic trajectories onto the problem of sampling binary sequences, making Monte Carlo methods directly applicable to the trajectory space \cite{Smith2022,DS26}.

\subsection{Importance sampling for the tails of $P_N(A)$}
It is generally not numerically feasible to use direct sampling to reach values of $A$ that fall within the tails of $P_N(A)$, as the probability decays very quickly. To access these rare events, we use a Metropolis importance sampling scheme on the bit sequence $\mathbf{b}$ along the lines of Ref.~\cite{Hartmann2025}.

In practice, we work with an extended trajectory of length $N + N_{\rm delay}$, where $N_{\rm delay} \simeq 0.1\, N$. The observable $A$ is computed only over the last $N$ steps, while the initial $N_{\rm delay}$ steps serve as a buffer that allows the Metropolis chain to freely adjust the endpoint $x_N$ of the trajectory. The number of single-bit flip proposals per sweep is set to $N_{\rm met} = N + N_{\rm delay}$, ensuring that on average every bit in the sequence is a candidate for flipping at each sweep.

We define a weight function $W(A) = A^{\nu}$, where $\nu$ is an exponent that concentrates sampling in the tail of interest. For the left tail ($A < \langle A \rangle$) one may set $\nu = 1$. For the right tail ($A > \langle A \rangle$), it is necessary to choose $\nu < 1/2$ to ensure convergence, since for large $A$ the distribution decays as $-\ln P_N(A) \sim \sqrt{A}$. The target distribution over bit sequences is proportional to $e^{k W(A(\mathbf{b}))}$, where $k$ biases the sampling toward large or small values of $A$. In our case, we chose $\nu = 1/4$.

Each Metropolis step proceeds as follows:
\begin{enumerate}
    \item Select a random index $n^{*}\in\left\{ 0,1,\ldots,N+N_{{\rm delay}}-1\right\} $ and flip bit $b_{n^*}$.
    \item Compute the new observable value $A_{\rm new}$ and the     acceptance ratio
    \begin{align}
        \alpha_{\rm acc} = \min\!\left(1,\, e^{k\left[W(A_{\rm new}) - W(A_{\rm cur})\right]}\right).
    \end{align}
    \item Accept the flip with probability $\alpha_{\rm acc}$, otherwise restore the original bit.
\end{enumerate}
After a burn-in period of $N_{\rm burn}$ sweeps, the chain explores
the region of trajectory space corresponding to the target value of
$A$, and samples collected during the production phase provide a
biased estimate
\begin{equation}
P_{N,k}\left(A\right)= Z\left(k\right)^{-1}P_{N}\left(A\right)e^{kW\left(A\right)}
\end{equation}  
of $P_N(A)$ near that value, where $Z(k)$ is a normalization factor. By running independent chains at different values of $k$, each targeting a different region of $A$, one obtains a set of overlapping biased distributions. These are then combined using a standard histogram reweighting procedure~\cite{Hartmann2025}, stitching the partial distributions together into a single estimate of $P_N(A)$ over a wide range of $A$. Multiple independent chains are run in parallel for each value of $k$, each seeded independently, and their outputs are concatenated before reweighting to improve statistics.

\section{Subleading-order corrections to Donsker-Varadhan theory and the power series of $\lambda(k)$ and $I(a)$}

\label{app:DVsubleading}

The goal of this Appendix is to obtain the subleading-order correction to the right tail of $P_N(A)$ in the subcritical (homogeneous) regime. We proceed in two steps. First, we derive a general subleading-order correction to the DV approximation, which is in fact valid whenever DV theory applies, including in the left tail. Second, we derive the relations between the coefficients of the power series of the rate function $I(a)$ and those of the SCGF $\lambda(k)$, which are needed in order to evaluate the correction in practice.

\subsection{Subleading-order correction to the DV approximation}
\label{app:DV-correction}

The rate function $I(A/N)$ is obtained from the SCGF $\lambda(k)$ via the
Legendre-Fenchel transform, as seen in Eq.~(\ref{eq:legendre-fenchel}). We are interested in the regime where $A$ is not much larger than the mean $\langle A \rangle$, i.e., where the rescaled variable $a \equiv A/N$ is close to $\lambda_1$, the first cumulant of $A/N$. In this regime, the Donsker-Varadhan (DV) approximation gives \cite{Touchette2009, Daniels1954, Bahadur1960}
\begin{align}
\label{subleadingDV}
    P_N(A) \simeq \frac{e^{- N \big[k^* a - \lambda(k^*)\big]}}{\sqrt{2 \pi N \lambda''(k^*)}} \, ,
\end{align}
where $k^* = k^*(a)$ is the saddle-point value of $k$, that is, the maximizer of $ka - \lambda(k)$.
Eq.~\eqref{subleadingDV} follows from an application of the subleading-order saddle-point approximation to the generating function of $P_N(A)$.
We are interested in the region where $I(a)$ and $\lambda(k)$ can be described by Taylor expansions, leading to the saddle-point condition
\begin{align}
    a = \lambda'(k^*) \,,
\end{align}
which is equivalent to the conjugate relation
\begin{align}
    k = I'(a) \,.
\end{align}
These two relations are the Legendre-transform correspondence between $\lambda$ and $I$, and can be combined into the functional identities
\begin{align}
    I'\left(\lambda'(k)\right) = k \,, \label{eq:legendre-identity}
\end{align}
and
\begin{align}
    \lambda'\!\bigl(I'(a)\bigr) = a \,.
    \label{eq:legendre-identity-2}
\end{align}
The term that appears in the exponential in Eq.~\eqref{subleadingDV} is simply $-N I(a)$, by the definition of the Legendre-Fenchel transform. In order to rewrite the pre-exponential factor in terms of the rate function as well, we differentiate both sides of Eq.~\eqref{eq:legendre-identity-2} with respect to $a$, to find
\begin{align}
    \lambda''\!\bigl(I'(a)\bigr)\, I''(a) = 1 \,.
\end{align}
Since the saddle-point condition requires $k^* = I'(a)$, this becomes
\begin{align}
    \lambda''\left(k^*\right) = \frac{1}{I''(a)} \,,
\end{align}
and substituting into Eq.~\eqref{subleadingDV} yields~\cite{Daniels1954, Bahadur1960}
\begin{align}
    P_N(A) \simeq \sqrt{\frac{I''(A/N)}{2\pi N}}\;
                   e^{-N I(A/N)} \,.
    \label{eq:DV-final}
\end{align}
This result is expected to hold generally in the regime where DV theory is valid. We expect it to hold also in the subcritical part of the right tail (i.e., for $\langle A \rangle < A < A_c$), where the rate function $I(a)$ is written as a power series and the result is interpreted as an asymptotic series. Eq.~\eqref{eq:DV-final} thus provides a subleading correction to Eq.~\eqref{eq:PNA-donsker-varadhan-extended}, and we verified its prediction by comparing to numerical simulations for $N=1600$ as shown in Fig.~\ref{fig:subleading} of the main text.

We take this opportunity to mention that the correction term $I''(A/N)$ in Eq.~\eqref{eq:DV-final} is expected to be important also in the case of sums of i.i.d. random variables, as a correction to Cram\'{e}r's theorem. This correction term could be useful to improve upon the recent results of Ref.~\cite{SmithCondensation25} (in which it was not taken into account).

\subsection{Relations between the power series of $\lambda(k)$ and $I(a)$}
\label{app:series-relations}

In order to use Eq.~\eqref{eq:DV-final} in practice, one must know the rate function $I(a)$, which we obtain as a power series around its minimum. The identity \eqref{eq:legendre-identity} allows one to systematically match the Taylor coefficients of $I(a)$ against those of $\lambda(k)$ \cite{CLDS26}. Writing the rate function as a series expansion centered at the mean $\lambda_1$ will greatly simplify the calculations. Explicitly, we have
\begin{align}
    I(a) &= I_2\,(a - \lambda_1)^2
         + I_3\,(a - \lambda_1)^3 \nonumber \\
         & \quad + I_4\,(a - \lambda_1)^4
         + O\!\left((a - \lambda_1)^5\right) \,.
    \label{eq:rate-series}
\end{align}
To extract the coefficients $I_2, I_3, I_4, \ldots$ in terms of the cumulants $\lambda_n \equiv \lambda^{(n)}(0)$ (see Appendix~\ref{app:numerical-SCGF-power-method}), we substitute the Taylor expansion
\begin{align}
    \lambda'(k) - \lambda_1 = \lambda_2\, k + \frac{\lambda_3}{2}\,k^2
                + \frac{\lambda_4}{6}\,k^3 + O(k^4)
\end{align}
into the identity~\eqref{eq:legendre-identity}, and expand the left-hand side
\begin{align}
    I'\!\bigl(\lambda'(k)\bigr)
        &= 2 I_2\bigl(\lambda'(k) - \lambda_1\bigr)
         + 3 I_3\bigl(\lambda'(k) - \lambda_1\bigr)^2 \nonumber \\
        & \quad + 4 I_4\bigl(\lambda'(k) - \lambda_1\bigr)^3
         + O\!\left(k^4\right) = k \,.
\end{align}
Matching powers of $k$ on both sides yields a system of equations that can be solved order by order:
\begin{align}
    O(k^1)&: ~
        2 I_2 \lambda_2 = 1 \quad\Longrightarrow\quad I_2 = \frac{1}{2\lambda_2} \,, \\
    O(k^2)&: ~
        2 I_2\,\frac{\lambda_3}{2} + 3 I_3\,\lambda_2^2 = 0
        \quad\Longrightarrow\quad
        I_3 = -\frac{\lambda_3}{6\lambda_2^3} \,, \\
    O(k^3)&: ~
        2 I_2\,\frac{\lambda_4}{6} + 3 I_3\,\lambda_2\lambda_3
        + 4 I_4\,\lambda_2^3 = 0 \nonumber \\
        & \quad\Longrightarrow\quad
        I_4 = \frac{\lambda_3^2}{8\lambda_2^5}
                   - \frac{\lambda_4}{24\lambda_2^4} \,.
\end{align}
Substituting back into Eq.~\eqref{eq:rate-series} yields the rate function
up to fourth order,
\begin{align}
    I(a) &= \frac{1}{2\lambda_2}\,(a - \lambda_1)^2
          - \frac{\lambda_3}{6\lambda_2^3}\,(a - \lambda_1)^3 \notag\\
         &\quad
          + \left(\frac{\lambda_3^2}{8\lambda_2^5}
                - \frac{\lambda_4}{24\lambda_2^4}\right)(a - \lambda_1)^4
          + O\!\left((a-\lambda_1)^5\right) \,.
    \label{eq:rate-series-fourth}
\end{align}
The leading term reproduces the Gaussian prediction in accordance with the central limit theorem. We used the method described above, and continued up to the 8th order, to calculate the coefficients $I_n$ for the doubling, tent and logistic maps from the $\lambda_n$'s which we had computed previously. The results are given in Table \ref{tab:rate-coefficients}.

\begin{table}[h]
\centering
\begin{tabular}{lccc}
\hline\hline
~ & Doubling & Tent & Logistic \\
\hline
$I_2$ & $0.1349351$  & $0.1841618$  & $5.066059 \times 10^{-2}$  \\
$I_3$ & $-0.1847774$ & $-0.4099663$ & $-6.186204 \times 10^{-2}$ \\
$I_4$ & $0.1016880$  & $0.5656745$  & $4.692523 \times 10^{-2}$  \\
$I_5$ & $-0.1433343$ & $-1.139397$                & $-5.168574 \times 10^{-2}$ \\
$I_6$ & $-0.2717794$ & $0.1399731$  & $3.759301 \times 10^{-3}$  \\
$I_7$ & $-1.629617$  & $-10.77921$  & $-0.1414242$ \\
$I_8$ & $-9.834040$  & $-70.11944$  & $-0.4593649$ \\
\hline\hline
\end{tabular}
\caption{Perturbative expansion coefficients of the rate function,
$I(a) = \sum_{n=2}^{8} I_n\,(a - \lambda_1)^n + O\!\left((a-\lambda_1)^9\right)$,
obtained from Table~\ref{tab:cumulants} via the recursive
Legendre-transform identity in Eq.~\eqref{eq:legendre-identity}.}
\label{tab:rate-coefficients}
\end{table}

\section{Condensed phase: Additional details and subleading corrections}
\label{app:condensation}

In this Appendix, we first compute the tail of the PDF of $A_1$ for a general class of one-dimensional maps to leading and subleading order. We then use this result to derive the approximation for $P_N(A)$ in the condensed phase, up to subleading order. This Appendix expands and adds more details to the analysis of Section~\ref{sec:condensation} of the main text, which generalizes the results beyond the leading order and includes more general maps.

\subsection{The PDF of $A_1$ for general maps \label{app-sub-PDF-A1}}

We now derive the probability of the condensed contribution to $A$ for general non-invertible maps $f:[0,1]\to[0,1]$, under the assumption that the map behaves near the origin as
\begin{align}
    f(x) \simeq \beta x \, , \quad x \ll 1 \, , \label{eq:trapping-linear-near-origin}
\end{align}
with $\beta > 1$. We further assume that the two preimages, $f^{-1}(x) = z_{1,2}$, of each point $x$ are equally likely under the invariant measure, that is,
\begin{equation}
    \frac{p_s(z_1)}{|f'(z_1)|} = \frac{p_s(z_2)}{|f'(z_2)|} = \frac{p_s(x)}{2} \, .
    \label{eq:equal-preimages}
\end{equation}
This property holds trivially for the doubling and tent maps, where $p_s = 1$ and $|f'| = 2$, and for the logistic map, where $p_s(z_\pm) = 2/(\pi\sqrt{x})$ and $|f'(z_\pm)| = 4\sqrt{1-x}$.

The extremely large values of $A$ are produced when a trajectory becomes ``trapped'' near the origin for many consecutive steps, contributing macroscopically to the observable. This trapping occurs as a consequence of Eq.~(\ref{eq:trapping-linear-near-origin}). We decompose the observable into two parts
\begin{align}
    A = A_1 + A_h \, ,
\end{align}
where $A_1$ is the contribution from the sequence of very small values
of $x$, and $A_h$ represents the remaining typical contribution. 
More precisely, we define $A_1$ using the centered observable,
\begin{align}
    A_1 = -\sum_{i=n}^{n+M}
    \bigl[\ln x_i - \langle \ln x \rangle_S\bigr] \, ,
\end{align}
so that $A_1$ measures only the excess contribution of the trapped sequence above its average. This ensures that $A_h$ is distributed closer to a typical sample of the full $N$-step observable, and thus justifies using the distribution in Eq.~(\ref{eq:PNA-donsker-varadhan-extended}) for $A_h$. 

\begin{figure}
    \centering
    \includegraphics[width=\linewidth]{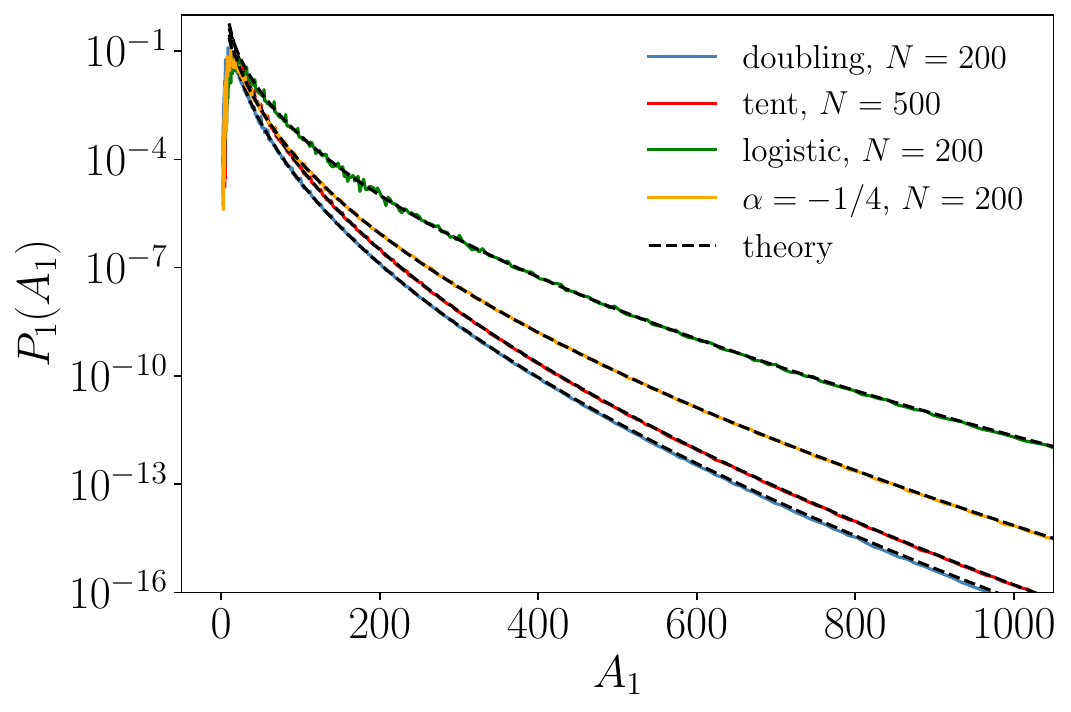}
    \caption{Comparison of the theoretical predictions for $P_1(A_1)$, given by Eqs.~\eqref{stretchedExponentialDoubling}, \eqref{stretchedExponentialLogistic} and \eqref{eq:P1_A1_alpha} (dashed lines), with the results of numerical simulations (solid lines).}
    \label{fig:A1-pdf-numerics-vs-theory}
\end{figure}

We now compute $A_1$ directly. Suppose a trajectory enters the very small $x$ region at step $n$, $x_n = \epsilon \ll 1$. The subsequent steps grow geometrically,
\begin{align}
    x_{n+m} \simeq \beta^m \epsilon \, ,
\end{align}
and the trajectory escapes the origin region after $M$ steps, defined
by imposing that the last element of the sequence is of order 1,
\begin{align}
    x_{n+M} = O(1) \longrightarrow M \simeq  -\frac{\ln \epsilon}{\ln \beta} \, .
    \label{eq:matching-P-x1}
\end{align}
The contribution of this sequence to the observable is
\begin{align}
    A_1 =& -\sum_{i=n}^{n+M}\left[\ln\left(\beta^{i-n}\epsilon\right)+\left\langle -\ln x\right\rangle _{S}\right]  \nonumber \\ 
    \simeq& - \frac{M(M+1)}{2}\ln\beta - (M+1) \ln \epsilon  \\
    - (M+1) \langle - \ln x \rangle_S \, , \nonumber
\end{align}
where we replace $M$ from Eq.~(\ref{eq:matching-P-x1}) to find
\begin{align}
    A_1 \simeq \frac{(\ln \epsilon)^2}{2\ln\beta} + \ln \epsilon \left(\frac{\langle - \ln x \rangle_S}{\ln \beta} - \frac{1}{2} \right) \, .
    \label{eq:A1-xn}
\end{align}
Eq.~\eqref{eq:A1-xn} is quadratic in $\ln\epsilon$, which we solve for $\ln\epsilon$ by completing the square and keeping the leading terms for large $A_1$,
\begin{align}
    \ln \epsilon \simeq -\sqrt{2\ln\beta\, A_1}
    + \left(\frac{\ln\beta}{2} - \langle{-\ln x}\rangle_S\right) \, .
\end{align}
Exponentiating, we obtain $\epsilon$ as a function of $A_1$,
\begin{align}
\label{epsilonOfA1}
    \epsilon \simeq \sqrt{\beta}\,e^{-\langle-\ln x\rangle_S}\,
    e^{-\sqrt{2\ln\beta\, A_1}} \, .
\end{align}
Since $A_1$ increases as $\epsilon \to 0$, a trajectory entering the
region $x_n < \epsilon$ necessarily produces a value of $A_1$ larger
than the value $\hat{A}_1(\epsilon)$ given by Eq.~(\ref{eq:A1-xn}).
The probability of finding $x$ in the range $[0,\epsilon]$ is given by the
invariant measure,
\begin{align}
    \mathrm{Prob}(0 < x < \epsilon) = \int_0^\epsilon p_S(x)\,dx
    \simeq \frac{\mu}{1+\alpha}\,\epsilon^{1+\alpha} \, ,
    \label{eq:CDF}
\end{align}
where $\alpha$ and $\mu$ characterise the behaviour of $p_S$ near the origin as $p_S(x) \simeq \mu x^\alpha$. For the doubling and tent maps, $\alpha=0$ and $\mu=1$, and for the logistic map, $\alpha=-1/2$ and $\mu=1/\pi$. Since entering the region $[0,\epsilon]$ is equivalent to producing a value $A_1 > \hat{A}_1(\epsilon)$, the tail probability of $A_1$ is obtained by summing this single-element probability over the $N$ possible positions of the minimum along the sequence, with an additional factor of $1/2$ accounting for the requirement that the preceding element lie on the larger of the two preimage branches, as explained in the main text [see the discussion preceding Eq.~\eqref{eq:prob-of-xmin-less-epsilon}]. Here and throughout, we assume that the two preimages of each point are equally likely, as is the case for all of the maps considered in this work. We thus obtain
\begin{align}
    \mathrm{Prob}(A_1 > \hat{A}_1) &\simeq \frac{N \mu}{2(1+\alpha)} \, \frac{\beta^{(1+\alpha)/2}}{e^{(1+\alpha)\langle-\ln x\rangle_S}} \nonumber\\
    &\quad \times\, e^{-(1+\alpha)\sqrt{2\ln\beta\,\hat{A}_1}} \, ,
\end{align}
where we have substituted the expression \eqref{epsilonOfA1} for $\epsilon$ as a function
of $\hat{A}_1$ and kept only the leading exponential contribution for
large $\hat{A}_1$. The factor $1/2$ relies on the equal-preimage property in Eq.\,\eqref{eq:equal-preimages}. The PDF of $A_1$ is then obtained by differentiation,
\begin{align}
    P_1(A_1) &= -\left.\partial_{\hat{A}_1}\,\mathrm{Prob}(A_1 > \hat{A}_1)\right|_{\hat{A}_1 = A_1} \nonumber \\
    &\simeq   \frac{N \mu \, \beta^{(1+\alpha)/2}}{2 e^{(1+\alpha)\langle-\ln x\rangle_S}} \sqrt{\frac{\ln \beta}{2 A_1}} e^{-(1+\alpha)\sqrt{2\ln\beta\,A_1}} \, , \nonumber \\
    \label{eq:PA1-app}
\end{align}
which is a one-sided stretched exponential with exponent $1/2$.
For the doubling and tent maps, $\alpha = 0$, $\beta = 2$ and $\langle - \ln x \rangle_S = 1$, giving
\begin{align}
\label{stretchedExponentialDoubling}
    P_1(A_1) \simeq \frac{N}{2e}
    \sqrt{\frac{\ln 2}{A_1}}\,e^{-\sqrt{ (2\ln 2)\,A_1}} \, ,
\end{align}
while for the logistic map, $\alpha = -1/2$, $\beta = 4$ and $\langle - \ln x \rangle_S = \ln 4$, giving
\begin{align}
\label{stretchedExponentialLogistic}
    P_1(A_1) \simeq \frac{N}{4 \pi}
    \sqrt{\frac{\ln 4}{A_1}}\,e^{-\sqrt{\ln 2\,A_1}} \, .
\end{align}
We compared these predictions with numerical computations of $P_1(A_1)$ for these 3 maps and found excellent agreement, see Fig.~\ref{fig:A1-pdf-numerics-vs-theory}.

To verify the validity of Eq.~(\ref{eq:PA1-app}) beyond the three main maps, we consider a fourth map built by an artificial conjugation of the logistic map by the power $x \mapsto x^\gamma$ with $\gamma = 2/3$. Explicitly, the forward map is $f(x) = [4x^{3/2}(1-x^{3/2})]^{2/3}$, which near the origin behaves as $f(x) \simeq \beta x$ with $\beta = 4^{2/3}$. The invariant measure follows from the change of variables $u = x^{3/2}$ applied to the logistic invariant measure for small $x$ behaves as $p_S(x) \simeq \frac{3}{2\pi}
x^{-1/4}$, so $\mu = 3/(2\pi)$, the exponent is $\alpha=-1/4$ and $\langle - \ln x \rangle_S = (4/3) \ln 2$. Replacing these values into Eq.~(\ref{eq:PA1-app}) gives 
\begin{align}
    P_1(A_1) \simeq
    \frac{N}{4\pi}
    \sqrt{\frac{\ln 8}{A_1}}\,
    e^{-\sqrt{\frac{3\ln 2}{2}\, A_1}} \, , \label{eq:P1_A1_alpha}
\end{align}
providing an additional test of the formula.

\subsection{Independence approximation and convolution structure}

In the appendix Sec.~\ref{app-sub-PDF-A1}, we argued that, for large $A > \langle A \rangle$, the observable is well approximated by the sum of two statistically independent contributions, $A \simeq A_1 + A_h$. Here $A_1$ is the condensed piece, i.e. the contribution from a relatively short run of $M \ll N$ consecutive small elements in the chaotic sequence, and $A_h$ is the contribution of the rest of the elements of the sequence, and follows more closely the probability of more typical events. We remind that $A_1$ measures only the excess contribution, above its average, ensuring that $A_h$ is distributed closer to the one in Eq.~(\ref{eq:DV-final}). Under this independence assumption, the distribution of the complete observable $A$ is the convolution
\begin{equation}
    P_N(A) \simeq \int_0^{A} d A_1\; P_1(A_1)\,P_h(A - A_1),
    \label{eq:convolution}
\end{equation}
where $P_1$ and $P_h$, which were derived in the previous sections, are found in Eqs.\,(\ref{eq:PA1-app})  and\,(\ref{eq:DV-final}) respectively. 
We have verified the validity of this expression numerically, meaning we numerically calculated the convolution with $P_1$ and $P_h$ obtained using the theoretical prediction, Eqs.\,(\ref{eq:PA1-app})  and\,(\ref{eq:DV-final}).\\

\subsection{Evaluation of the convolution \label{app-sub-convolution-evaluation}}

We now have at hand both ingredients of the convolution in Eq.~\eqref{eq:convolution}, namely, the distribution $P_1$ of the condensed contribution, given by Eq.~\eqref{eq:PA1-app}, and the distribution $P_h$ of the homogeneous contribution, given by Eq.~\eqref{eq:DV-final}. We can proceed to evaluate $P_N(A)$, both numerically and theoretically.

At moderate values of $N$, we evaluate the integral in Eq.~\eqref{eq:convolution} numerically, using the theoretical expressions for the two distributions, where for $P_h$ we truncate the perturbative expansion of the rate function in Eq.~ \eqref{eq:rate-series-fourth} at the fourth order, and for $P_1$ we retain the pre-exponential factor, as in Eq.~\eqref{eq:PA1-app}. In practice, the integration over $A_h$ is restricted to the window $\left\langle A\right\rangle \pm 6\sqrt{\lambda_2 N}$, that is, six standard deviations of the typical fluctuations. For the values of $A$ shown in Fig.~\ref{fig:subleading} this amounts to a cutoff at small $A_1$, which excludes the region where the asymptotic form \eqref{eq:PA1-app} does not apply. The accuracy of the convolution improves as $A$ increases, since the dominant values of $A_1$ move deeper into the asymptotic regime, and beyond $A_c + 2 N^{2/3}$ it coincides with the Laplace evaluation presented below. The resulting prediction is compared with importance-sampling simulations in Fig.\,\ref{fig:subleading} of the main text, with excellent agreement even at moderate values of $N$, provided that $A$ is sufficiently far into the right tail for the asymptotic form \eqref{eq:PA1-app} to apply.

For $N$ sufficiently large, the integral can instead be evaluated analytically using Laplace's method, as we now show. In the transition regime, where $A - \left\langle A\right\rangle \sim N^{2/3}$, we employ the scaled variables of Sec.\,\ref{sec:condensation}, $A - \left\langle A\right\rangle = N^{2/3} y$ and $A_1 = N^{2/3} z$, in terms of which the argument of the rate function in Eq.~\eqref{eq:DV-final} becomes $(A - A_1)/N = \lambda_1 + N^{-1/3}\left(y - z\right)$, where we used $\left\langle A\right\rangle = N \lambda_1$. Plugging Eqs.~\eqref{eq:PA1-app} and \eqref{eq:DV-final} into Eq.~\eqref{eq:convolution}, we can isolate the leading exponential contribution of the integrand as $e^{-\Phi(z)}$, where
\begin{equation}
    \Phi(z) \equiv N\,I\!\left(\lambda_1 + \frac{y - z}{N^{1/3}}\right) + c\,N^{1/3}\sqrt{z}
    \label{eq:exponent}
\end{equation}
and
\begin{equation}
    c = (1+\alpha)\sqrt{2\ln\beta} \, .
    \label{eq:c-general}
\end{equation}
For the doubling and tent maps, Eq.~\eqref{eq:c-general} gives $c = \sqrt{2\ln 2}$, as we see in the main text, while for the logistic map it gives $c = \sqrt{\ln 2}$. In the transition regime, the argument of the rate function in Eq.~\eqref{eq:exponent} deviates from its minimum $\lambda_1$ only by $O(N^{-1/3})$, so that for sufficiently large $N$, the rate function $I(a)$ may be replaced by its parabolic approximation $I(a) \simeq I_2\left(a - \lambda_1\right)^2$, and the exponent takes the scaling form
\begin{equation}
    \Phi(z) \simeq N^{1/3}\,\mathcal{F}(y,z) \, ,
    \label{eq:Phi-scaling}
\end{equation}
where $\mathcal{F}(y,z) = c\sqrt{z} + I_{2}\left(y-z\right)^{2}$ is the scaling function in Eq.~\eqref{Fdef}. In practice, this minimization is easily carried out numerically. For a given value of $y$, one locates the global minimizer $z^*$ of $\mathcal{F}(y,z)$ on the interval, and the Laplace approximation of the convolution \eqref{eq:convolution} then reads
\begin{equation}
    P_N(A) \simeq P_1\!\left(A_1^*\right) P_h\!\left(A - A_1^*\right)
    \sqrt{\frac{2\pi N}{\partial_z^2\,\mathcal{F}\left(y, z^*\right)}} \, ,
    \label{eq:laplace-general}
\end{equation}
where $A_1^* = N^{2/3} z^*$ is the condensed contribution at the minimizer, $\partial_z^2 \mathcal{F} = 2 I_2 - c/(4 z^{3/2})$, and the distributions $P_1$ and $P_h$, including their pre-exponential factors, are given by Eqs.~\eqref{eq:PA1-app} and \eqref{eq:DV-final} respectively. The remainder of this subsection is devoted to carrying out the minimization analytically.

In the limit $N \to \infty$, the integral is therefore dominated by the neighborhood of the global minimum of $\mathcal{F}(y,z)$ over $z$ at fixed $y$. We now analyze the structure of the minima of $\mathcal{F}$ in detail.

The derivative $\partial_z \mathcal{F} = c/(2\sqrt{z}) - 2 I_2 \left(y - z\right)$ diverges to $+\infty$ as $z \to 0^+$, so the boundary point $z = 0$, which describes the homogeneous phase, is always a local minimum of $\mathcal{F}$. In particular, for $A$ near $\left\langle A\right\rangle$, that is, for small $y$, it is the only minimum. By requiring $\partial_z \mathcal{F} = 0$, we find that interior stationary points satisfy
\begin{equation}
    4 I_2 \left(y - z\right) \sqrt{z} = c \, ,
    \label{eq:saddle}
\end{equation}
which is equivalent to balancing the slope of the rate function against the slope of the stretched-exponential tail of $P_1$. The left-hand side of Eq.~\eqref{eq:saddle} vanishes at both endpoints of the interval $z \in [0,y]$, and attains its maximum $8 I_2 y^{3/2} / (3\sqrt{3})$ at $z = y/3$. Therefore, we have three distinct cases: 1) no stationary points when $c$ is bigger than the maximum, 2) two points when $c$ is smaller than the maximum, and 3) one single point, when the maximum is exactly $c$. The latter happens at $y \geq y_{\mathrm{sp}}$, where
\begin{equation}
    y_{\mathrm{sp}} = \frac{3}{4}\left(\frac{c}{I_2}\right)^{2/3} ,
    \label{eq:ysp}
\end{equation}
corresponding to $A_{\mathrm{sp}} = \left\langle A\right\rangle + y_{\mathrm{sp}} N^{2/3}$. For $y > y_\mathrm{sp}$, we have a local minimum to compete with $z=0$. At $y = y_{\mathrm{sp}}$, a single degenerate stationary point appears at $z = y_{\mathrm{sp}}/3$, and for $y > y_{\mathrm{sp}}$ it splits into two solutions $z_- < y/3 < z_+$. The nature of the two stationary points follows from the curvature $\partial_z^2 \mathcal{F} = 2 I_2 - c/(4 z^{3/2})$, which, upon eliminating $c$ using Eq.\,\eqref{eq:saddle}, becomes
\begin{equation}
    \partial_z^2 \mathcal{F} \,\big|_{\partial_z \mathcal{F} = 0} = I_2 \, \frac{3z - y}{z} \, .
    \label{eq:curvature}
\end{equation}
The curvature is negative at $z_- < y/3$ and positive at $z_+ > y/3$, so $z_-$ is a local maximum of $\mathcal{F}$, while $z_+$ is a local minimum, describing the condensed phase.

For $y > y_{\mathrm{sp}}$, the global minimum of $\mathcal{F}$ is determined by the competition between the boundary value $\mathcal{F}(y,0) = I_2 y^2$ and the interior value $\mathcal{F}(y, z_+)$. At the spinodal, substituting $c = 8 I_2 y_{\mathrm{sp}}^{3/2} / (3\sqrt{3})$ gives $\mathcal{F}(y_{\mathrm{sp}}, y_{\mathrm{sp}}/3) = \frac{4}{3} I_2 y_{\mathrm{sp}}^2 > \mathcal{F}(y_{\mathrm{sp}}, 0)$, so the homogeneous one still dominates. As $y$ grows, however, the contribution from the interior minimum becomes larger. The two branches cross exactly once, at the critical value $y_c$ defined by the condition $\mathcal{F}\left(y_c, z_+(y_c)\right) = I_2 y_c^2$. This condition may be rewritten as $c = I_2 \sqrt{z}\left(2y - z\right)$, and equating with \eqref{eq:saddle} gives $4\left(y - z\right) = 2y - z$, so the minimizer at the transition is $z_c = 2 y_c / 3$. Substituting back into \eqref{eq:saddle} then yields
\begin{equation}
    y_{c} = \frac{3}{2^{5/3}}\left(\frac{c}{I_{2}}\right)^{2/3} , \qquad
    z_{c} = \frac{2 y_{c}}{3} = \left(\frac{c}{2 I_{2}}\right)^{2/3} ,
    \label{eq:yc-zc}
\end{equation}
the former reproducing the value of $y_c$ given in the main text, and corresponding to the critical value of the observable $A_c$, explicitly
\begin{equation}
    \label{eq:Ac-deff}
    A_c \simeq \left\langle A\right\rangle + y_c N^{2/3} = \left\langle A\right\rangle +  \frac{3}{2}\left(\frac{c N}{2 I_{2}}\right)^{2/3} \, .
\end{equation}
Comparing with \eqref{eq:ysp}, we find $y_{\mathrm{sp}} = 2^{-1/3}\, y_c$, so the condensed minimum is formed only shortly before it becomes dominant. When we cross $y = y_c$, the global minimizer of $\mathcal{F}$ thus jumps discontinuously from the boundary $z = 0$ to the interior point $z_c$. The jump of the minimizer is the mechanism behind the first-order DPT discussed in the main text.

Beyond the leading exponential order, the Laplace evaluation of the convolution also yields the pre-exponential factors. At finite $N$, we return to the full exponent $\Phi(z)$ of Eq.~\eqref{eq:exponent}, in which the rate function is not replaced by its parabolic approximation. For $A > A_c$, the integral in Eq.~\eqref{eq:convolution} is dominated by the neighborhood of the interior minimizer $z^*$ of $\Phi(z)$, which satisfies
\begin{equation}
    I'\!\left(a^*\right) = \frac{c}{2 N^{1/3} \sqrt{z^*}} \, , \qquad
    a^* \equiv \lambda_1 + \frac{y - z^*}{N^{1/3}} \, ,
    \label{eq:saddle-finite-N}
\end{equation}
and which reduces to the solution $z_+$ of Eq.~\eqref{eq:saddle} when $I'(a)$ is replaced by $2 I_2 (a - \lambda_1)$. Expanding $\Phi$ to second order around $z^*$, we find the curvature
\begin{equation}
    \partial_z^2 \Phi\left(z^*\right) = N^{1/3}\left[I''\!\left(a^*\right) - \frac{c}{4 \left(z^*\right)^{3/2}}\right] ,
    \label{eq:curvature-finite-N}
\end{equation}
so that the Gaussian integration over $A_1 = N^{2/3} z$ contributes a factor $\sqrt{2\pi N^{4/3} / \partial_z^2 \Phi(z^*)}$. Evaluating the pre-exponential factors of Eqs.\,\eqref{eq:PA1-app} and \eqref{eq:DV-final} at $A_1^* = N^{2/3} z^*$ and at $a^*$, respectively, the powers of $N$ combine into an overall $N^{2/3}$, and we obtain
\begin{eqnarray}
    P_N(A) &\simeq&
    \frac{N^{2/3} \mu \, \beta^{(1+\alpha)/2}}{2\, e^{(1+\alpha)\langle-\ln x\rangle_S}} \sqrt{\frac{\ln \beta}{2 z^*}} \nonumber\\
    & \times & \left[\frac{I''(a^*)}{I''(a^*) - \tfrac{c}{4 (z^*)^{3/2}}}\right]^{1/2}
    e^{-\Phi(z^*)} \, ,
    \label{eq:laplace-subleading}
\end{eqnarray}
Eq.~\eqref{eq:laplace-subleading} retains the higher-order terms of $I(a)$, through $\Phi(z^*)$, $a^*$ and $I''(a^*)$, as well as the pre-exponential factors. It reduces to Eq.~\eqref{eq:laplace-general} when the rate function is truncated at quadratic order, in which case $I''(a^*) = 2 I_2$ and $z^* = z_+$.

Finally, we show that deep in the condensed phase, the complete expression \eqref{eq:laplace-subleading} reduces to $P_1$ evaluated at $A - \left\langle A\right\rangle$, including the pre-exponential factor. Restoring the unscaled saddle point $A_1^* = N^{2/3} z^*$ and writing $\Phi(z^*) = N I(a^*) + c\sqrt{A_1^*}$, the pre-exponential factor of Eq.~\eqref{eq:laplace-subleading} coincides with that of Eq.~\eqref{eq:PA1-app} evaluated at $A_1^*$, so that Eq.~\eqref{eq:laplace-subleading} may be rewritten as
\begin{equation}
    P_N(A) \simeq P_1(A_1^*) \,
    \left[\frac{I''(a^*)}{I''(a^*) - \tfrac{c}{4 (z^*)^{3/2}}}\right]^{1/2}
    e^{-N I(a^*)} \, .
    \label{eq:bigjump-reduction}
\end{equation}
Deep in the condensed phase, the saddle-point condition \eqref{eq:saddle-finite-N} gives $z^* \approx y$, so $a^* \to \lambda_1$ and $A_1^* \to A - \left\langle A\right\rangle$, while the two correction factors in Eq.~\eqref{eq:bigjump-reduction} tend to one, since $N I(a^*) \simeq c^2 N / (16 I_2 A_1^*) \to 0$ and $c/(z^*)^{3/2} \to 0$. In this limit, which is reached for $A - \left\langle A\right\rangle \gg N$, we obtain
\begin{equation}
    P_N(A) \simeq P_1\!\left(A - \left\langle A\right\rangle\right) ,
    \label{eq:bigjump-limit}
\end{equation}
in agreement with the leading-order result \eqref{PNABigJump} of the main text, and now including the pre-exponential factor. This is to be expected, as in this regime the entire fluctuation is carried by the big jump, while the homogeneous contribution remains typical, so that the normalized distribution $P_h$ merely integrates to unity in the convolution in Eq.~\eqref{eq:convolution}.

\end{document}